\documentclass[twoside,twocolumn,9pt]{revtex4-1}
\usepackage{graphicx}
\usepackage{amsmath}
\usepackage{amssymb}

\begin{document}

\title{A soft departure from jamming: the compaction of deformable granular matter under high pressures} 

\author{Joel T. Clemmer}
\affiliation{Sandia National Laboratories, Albuquerque, New Mexico 87185, USA}
\author{Joseph M. Monti}
\affiliation{Sandia National Laboratories, Albuquerque, New Mexico 87185, USA}
\author{Jeremy B. Lechman}
\affiliation{Sandia National Laboratories, Albuquerque, New Mexico 87185, USA}
\date{\today}

\begin{abstract}

The high-pressure compaction of three dimensional granular packings is simulated using a bonded particle model (BPM) to capture linear elastic deformation.
In the model, grains are represented by a collection of point particles connected by bonds.
A simple multibody interaction is introduced to control Poisson's ratio and the arrangement of particles on the surface of a grain is varied to model both high- and low-frictional grains.
At low pressures, the growth in packing fraction and coordination number follow the expected behavior near jamming and exhibit friction dependence.
As the pressure increases, deviations from the low-pressure power-law scaling emerge after the packing fraction grows by approximately 0.1 and results from simulations with different friction coefficients converge.
These results are compared to predictions from traditional discrete element method simulations which, depending on the definition of packing fraction and coordination number, may only differ by a factor of two. 
As grains deform under compaction, the average volumetric strain and asphericity, a measure of the change in the shape of grains, are found to grow as power laws and depend heavily on the Poisson's ratio of the constituent solid. 
Larger Poisson's ratios are associated with less volumetric strain and more asphericity and the apparent power-law exponent of the asphericity may vary.
The elastic properties of the packed grains are also calculated as a function of packing fraction.
In particular, we find the Poisson's ratio near jamming is 1/2 but decreases to 1/4 before rising again as systems densify.

\end{abstract}

\maketitle

%==============================================================%
\section{Introduction}
\label{sec:intro}
%==============================================================%

Across industry and nature, dense granular matter experiences a diverse range of environments. 
Such environments are often representative of the hard-particle limit where grains experience stresses much less than their elastic moduli causing minimal elastic deformation \cite{Torquato2010}.
For instance, this is frequently the case for relatively stiff materials like sand or rocks under the force of gravity in stationary piles, chute flows, hoppers, etc.
However, under relatively high-pressure conditions, contact forces can induce significant particle deformation and the hard-particle approximation is not accurate.
This is seen in applications including pharmaceutical die compaction of powders, rock cataclasis due to fault motion, and impacts on foam packaging material or rubber mulch in playgrounds.
Such high loads can lead to significant elastic deformation, plasticity, and fracture in grains, altering the mechanical properties of the granular medium.

At low pressures, granular packings are often characterized in the context of the jamming transition \cite{VanHecke2010, Behringer2019}.
Granular materials transition to a rigid state that can hold a pressure when the packing fraction increases to a critical value $\phi_c$.
As $\phi$ continues to increase, the pressure, as well as other quantities, grows as a power of $\phi-\phi_c$. 
In granular materials, the specific value of $\phi_c$ depends on friction coefficients, but power laws are found to be independent of friction \cite{Silbert2010, Santos2020}.

At high pressures, the behavior of granular materials has been less explored in the literature.
Experimentally, various techniques have been used to probe highly compressed granular media \cite{Erikson2002, Brodu2015, Vu2020, Bares2022} finding notable changes in behavior including a transformation in the functional shape of distributions of contact forces \cite{Erikson2002} and a transition to continuum behavior \cite{Bares2023} with increasing pressure.
Computational studies have leveraged both finite element method (FEM) based models \cite{Gethin2003, Procopio2005, Cantor2020, Cardenas-Barrantes2022}, the material point method (MPM) \cite{Nezamabadi2017, Vu2020, Vu2021}, and bonded particle models (BPM) \cite{Nezamabadi2017, Dosta2017, Giannis2023} to explicitly represent the elastic deformation of grains.
While most simulations have either been limited to two dimensions or only explored comparisons in compaction curves to experiments, work by \citet{Cardenas-Barrantes2022} in particular used the FEM-based Non-Smoothed  Contact Dynamic Method (NSCD) to quantify the scaling of a wide range of metrics, including the pressure, coordination number, and the asphericity of grains, with the distance from the jamming transition in this limit.
A related set of systems include emulsions, foams, and biological cells which undergo significant deformation but do not necessarily exhibit internal solid elasticity like grains.
The particles in these systems can be more accurately viewed as elastic, fluid-filled membranes that still penalize changes in volume or surface area, however, similar challenges exist between the two classes of problems implying research in either direction is mutually beneficial.
Recent work using the Deformable Particle Method (DPM) has significantly progressed computational methods to model these systems and has characterized many relevant features of high-pressure packings in both two \cite{Boromand2018, Boromand2019} and three dimensions \cite{Wang2021}.
Although solid elasticity may not be explicitly modeled, results are likely still very applicable to the problem.

In this work, we focus on the isotropic linear elastic response of deformable grains under compaction, or what has been termed squishy granular material \cite{Bares2022}.
In addition to being the primary regime of interest in materials which can sustain significant reversible strains, elastic deformation always serves as a precursor to potential plastic deformation or fracture and is therefore an important limit to establish before overlaying such additional inelastic mechanisms or extending to nonlinear elasticity.
The goals of this article are twofold.
The first is to develop a computationally efficient and flexible method that can represent the large deformations of an ensemble of grains.
Computational efficiency is necessary to simulate a sufficiently large number of grains to avoid significant finite size effects associated with jamming \cite{Goodrich2015} and minimize fluctuations in mechanical responses.
For instance in shear jamming, there are significant fluctuations in systems with $<10^3$ grains \cite{Santos2022}.
Flexibility is necessary to model granular media with a breadth of material models, including both linear and nonlinear elasticity, surface friction, adhesion and cohesion, and plasticity and fraction. 
It is also desirable to have a grain model that readily resolves aspherical shapes or inhomogeneous materials. 
To accomplish this, we propose a BPM that includes a new multibody term to control isotropic linear elasticity and developed an open source BPM package in the parallelized particle simulator LAMMPS \cite{Thompson2021}.
We illustrate how friction can be adjusted in a BPM by controlling the morphology of a grain's surface.

The other primary goal of this paper is to characterize the scaling of standard features of jamming at large pressures of a packing comprised of isotropic, linear elastic grains.
Here we use BPM to model a wide range of elastic and frictional grains to characterize the impact of material properties on deformation. 
Results are compared to those from traditional discrete element method (DEM) simulations to highlight where resolution of internal degrees of freedom, which DEM lacks, is necessary. 
We find that if one considers certain metrics, results from DEM simulations only significantly deviate when the packing fraction $\phi$ exceeds $\phi_c$ by 0.1 and, even then, are within a factor of two from results from BPM simulations.
In addition, we describe how Poisson's ratio affects the deformation of grains.
Both the average volumetric strain of grains and their average distortion from their initial spherical shape, or asphericity, grow approximately as a power of the distance to jamming, $\phi - \phi_c$. 
At higher Poisson's ratios, grains exhibit less volumetric strain but more asphericity, the latter of which interestingly may have an exponent that depends on Poisson's ratio.
Lastly, we also characterize the elastic moduli of the packed systems.
Both the bulk and shear moduli initially grow as a power of excess packing fraction up as the system densifies until $\phi - \phi_c \sim 0.1$ before accelerating.
Interestingly, the Poisson's ratio of the packing $\nu_P$ is minimized at this transition with $\nu_P \sim 1/4$.

%==============================================================%
\section{Methods}
\label{sec:methods}
%==============================================================%

In traditional DEM simulations, each grain is represented as a single computational particle with translational and rotational degrees of freedom \cite{Cundall1979}.
Particles exchange contact forces with neighbors and trajectories are numerically integrated.
Typically, DEM simulations treat grains as spherical objects with Hertzian normal forces.
DEM is generally used in the hard-particle limit, although, there are more sophisticated contact models that account for non-linear elasticity or multicontact interactions to improve accuracy at high pressures \cite{Gonzalez2012}.
However, to explicitly model elastic deformation one inevitably needs to add internal degrees of freedom to grains to describe the internal strain field.
One solution is to solve the internal solid mechanics of a grain using the Finite-Element Method as done in both the Multi-Particle Finite Element Method (MPFEM) \cite{Gethin2003, Procopio2005, Harthong2012} and the NSCD \cite{Cantor2020, Cardenas-Barrantes2022}.
However, meshed-based methods can suffer from mesh entanglements at severe deformations, which may become relevant for complex grain geometries, and often struggle to represent other discontinuous behavior relevant to compacted grains such as fragmentation.
Another option is to use mesh-free formulations of continuum elasticity such as peridynamics \cite{Behzadinasab2018, Silling2021} or the material point method \cite{Homel2017, Nezamabadi2017, Vu2020, Vu2021}.
However, these methods can be fairly complex and computationally expensive.

In contrast to the above continuum methods, another popular approach are BPMs \cite{Lisjak2014}. 
In a BPM, grains are represented by a collection of particles connected by a predefined network of bonds.
The relative displacement of particles represents internal strain which then incurs forces or stresses from the bonds.
While elasticity naturally emerges from this framework, one has to design the forces exchanged by bonds to create the desired mechanical response as opposed to directly inputting a constitutive equation.
The obvious downside is that it is not always known apriori how to design or calibrate bond forces.
Unlike FEM, there is also little information on convergence to analytic continuum mechanical solutions with increasing simulation resolution.
However, a benefit of BPMs is their relatively simple formulation which provides substantial flexibility and computational efficiency.

For this work, we use a BPM approach and construct a new bond formulation to model isotropic, linear elastic systems.
In Sec. \ref{sec:model}, we describe this model and demonstrate its ability to overcome restrictions on Poisson's ratio that typically impair BPMs in Subsection \ref{sec:calibrate}.
We then describe the construction of spherical grains, the verification that the model reproduces Hertzian contact forces, and the creation of initial jammed states in Sec. \ref{sec:simultions}.
For comparison, we identify several parameterizations of traditional DEM simulations which approximately match the behavior of BPM simulations at low pressures in Sec. \ref{sec:dem}.
Lastly in Sec. \ref{sec:lammps}, we briefly describe the development of a BPM package in the open-source Large Scale Atomic/Molecular Massively Parallel Simulator (LAMMPS) codebase \cite{Thompson2021}.

%==============================================================%
\subsection{Bonded particle models}
\label{sec:bpm}
%==============================================================%

To the authors' knowledge, there is no consensus on the definition of a bonded particle model or BPM.
Many different computational models have been labeled BPMs \cite{Potyondy2004, Lisjak2014} and there are many different names for related models based on bonded interactions between particles including, but not limited to, the cohesive beam model \cite{Andre2012}, bonded DEM \cite{Celigueta2017}, and various lattice or spring network models \cite{Ostoja-Starzewski2002, Cusatis2011, Zhao2011, Chen2014, Chen2016, Kot2017, Golec2020}.
Additionally, the relation between BPMs and bond-based peridynamics (in contrast to the more complex state-based peridynamics) is not always clear and the two methods can share strengths and limitations \cite{Silling2010, Trageser2020}. 
In this article, we use the term bonded particle model or BPM to loosely refer to any particle-based method that attempts to model solid elasticity by exchanging forces between neighboring particles using a predefined bond network and an unambiguous stress-free reference state. 

Within BPMs, one can begin to break down different implementations based on numerical details.
For instance, particles that make up a solid body may be aligned on a lattice or may have a disordered configuration.
Lattices greatly simplify the calibration \cite{Wang2009b}, but are not ideal for representing isotropic materials as they can lead to anisotropic artifacts in crack propagation or contact forces.
To avoid such issues, we focus on disordered arrangements of particles.
Another important distinction is the type of particles used.
While there are some models which use aspherical particles \cite{Timar2011, Andre2019}, models typically use either spherical particles with rotational degrees of freedom or point particles with no rotational degrees of freedom.
With rotation, simulations are more akin to traditional DEM and bonds between particles can be thought of as beams which transmit normal and shear forces as well as torques \cite{Potyondy2004,Carmona2008,Wang2009,Andre2012}.
In contrast with point particles, bonds typically only transmit normal forces \cite{Beale1988, Chen2014, Chen2016, Kot2017, Clemmer2022}, although, additional forces may be overlaid \cite{Kirkwood1939, Schwartz1985, Ostoja-Starzewski2002, Zhao2011, Reid2018, Clemmer2023b}.

A common obstacle in BPMs is modeling different Poisson's ratios.
In a disordered packing of particles which only exert pairwise, central-body forces, Poisson's ratio $\nu$ is restricted to 1/3 in 2D and 1/4 in 3D \cite{Walton1987, Greaves2013}.
This restriction, part of what is known as Cauchy's relations, is partially circumvented in BPMs with rotational degrees of freedom as beam-like bonds also exert tangential forces such that increasing the strength of tangential forces relative to normal forces forces increases the relative resistance to shear.
This generally increases the ratio of the shear modulus to the bulk modulus which decreases Poisson's ratio \cite{Andre2012, Leclerc2019, Nguyen2019}. 
This effect has been extended in the deformable DEM model (DDEM) \cite{Rojek2021} where particles deform into ellipsoids based on their local stress state.
A similar increase in shear strength is also achieved in some point-particle-based BPMs by constructing additional force terms such as a rotationally-invariant tangential force in the Distinct Lattice Spring Model (DLSM) \cite{Zhao2011} or three-body angular interactions \cite{Kirkwood1939, Schwartz1985, Ostoja-Starzewski2002, Reid2018, Clemmer2023b}.
However, simulating larger Poisson's ratios can still prove difficult since reducing the shear modulus relative to the bulk modulus requires negative tangential or angular stiffnesses which could reduce stability.  

Another approach is to add a nonlocal or multibody term that depends on other nearby particles or bonds.
For instance, a nonlocal energetic term is constructed from the displacements of first and second neighbor bonds on a regular lattice in the Lattice Particle Model  \cite{Chen2014, Chen2016}, a quadratic energy term penalizes local volume changes in the Hybrid Mass Spring System \cite{Golec2020}, and a unique dispersion of incoming forces from particles onto their neighbors is used in the Extended Mass Spring Model \cite{Kot2017}.
In all of these models, one can tune the strength of the additional term to to model arbitrary Poisson's ratios, both increasing and decreasing the resistance of the solid to shear relative to compression.
Lastly, the mechanisms in DDEM \cite{Rojek2021} and  DLSM \cite{Zhao2011} described above could also be classified as multibody interactions since interactions depend on adjacent particles.

%==============================================================%
\subsection{Multibody bond formulation}
\label{sec:model}
%==============================================================%

In this work, the motivation is to find a simple BPM formulation that can represent Poisson's ratio $\nu$ both above and below $1/4$ in 3D while limiting computational costs, avoiding assumptions about the underlying particle arrangement, obeying physical symmetries, and conserving momentum. 
To avoid the computational costs associated with rotational degrees of freedom, we use point particles.
Bonds between particles $i$ and $j$ exert central-body forces with a magnitude given by
\begin{equation}
k_B \left(r_{0,ij} - r_{ij} \right) + 
a_B \left( \left[ \frac{V_i + V_j}{V_{0,i} + V_{0,j}} \right]^{1/3} - 
                  \frac{r_{ij}}{r_{0,ij}} \right)
\label{eq:fbond}                      
\end{equation}
where $k_B$ and $a_B$ are constants, $r_{ij}$ and $r_{0,ij}$ are the current and reference distances between particles, and $V_i$ and $V_{0,i}$ are measures of the current and reference local volumes occupied by each particle.
The first term in Eq.~\eqref{eq:fbond} proportional to $k_B$ represents a simple spring. 
On its own, this term could be used to represent a linear elastic material with $\nu = 1/4$ in 3D and a bulk modulus $K$ that depends on $k_B$ and the specific bond topology (e.g. how many bonds a particle has on average).
The second term proportional to $a_B$ is constructed to resemble a deviatoric term, the difference between the local volumetric dilation $D_{ij} \equiv \left[(V_i + V_j)/(V_{0,i} + V_{0,j})\right]^{1/3}$ and the stretch of a bond $\lambda_{ij} \equiv r_{ij}/r_{0,ij}$, and controls the shear modulus $G$ while having a minimal impact on the bulk modulus $K$.
Note that forces between bonded particles are still equal-and-opposite.

\begin{figure}
\begin{centering}
	\includegraphics[width=0.65\columnwidth]{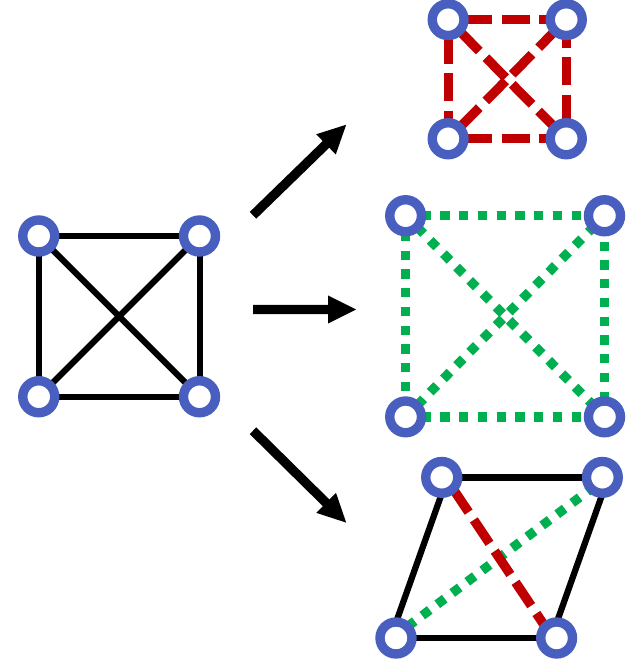}
	\caption{An idealized bond network between four particles undergoing isotropic contraction (top), isotropic extension (middle), and simple shear (bottom). Red, dashed bonds contract and green, dotted bonds expand to linear order in strain.}
	\label{fig:bond_demo}
\end{centering}
\end{figure}

To better conceptualize the multibody term, one can consider a few idealized deformation geometries, illustrated in Fig.~\ref{fig:bond_demo}, to linear order in strain.
Under pure isotropic compression or extension, all bonds stretch by a factor $\lambda$ regardless of orientation.
As all particles dilate by a factor $D = \lambda$, the second term then evaluates to zero.
The stress state depends on the first harmonic term so $K$ is minimally dependent on $a_B$ (in disordered systems there is some dependence which is further discussed in the following section).

In simple shear, one bond is extended while another is contracted (to linear order) in Fig. \ref{fig:bond_demo}.
Therefore, two particles experience a reduction in local volume while the other two particles experience an equal and opposite increase.
For the four exterior bonds, these two effects cancel out and the multibody term is again irrelevant.
However, the average volumetric dilation and the bond stretch are less than unity in the internal compressed bonds. 
The bond stretch is larger in magnitude than the volumetric dilation such that their difference is positive.
For the internal extended bond, the opposite is true such that the difference is negative. 
Thus, forces and the shear stress increase if $a_B > 0$ and decreases if $a_B < 0$.

Before calculating forces, the local volume $V_i$ is first calculated for each particle $i$.
Instead of calculating a geometrically exact volume, we use a proxy
\begin{equation}
V_i = \sum_{j\ne i} r_{ij}^3
\end{equation}
where the sum occurs over neighboring bonded particles.
Since we evaluate ratios of volumes, constant prefactors are neglected and bonded neighbors are assumed to be evenly distributed on a spherical surface.
While a more accurate metric might be ideal, we find that this approximation performs well while minimizing computational costs.
While exact costs depend on implementations, we find this multibody formulation approximately doubles the computational time needed to calculate forces. 
In contrast, we find forces that account for rotational degrees of freedom can take approximately ten times longer to calculate than a simple harmonic term\footnote{As seen in examples of a brittle plate impact included in the main LAMMPS distribution}.

In addition to bond forces described in Eq.~\eqref{eq:fbond}, the model also includes damping forces and non-bonded interactions between neighboring particles.
Between bonded particles, there is an additional dissipative, central-body force with magnitude
\begin{equation}
-\Gamma \hat{r}_{ij} \cdot \delta \vec{v}_{ij}
\end{equation}
where $\delta \vec{v}_{ij}$ is the difference in velocity between particles.
This term originates from Dissipative Particle Dynamics and damps the difference in normal velocities while conserving linear and angular momentum \cite{Groot1997}.
Between non-bonded particles, such as particles on the surfaces of separate grains, an alternative central-body force is applied between particles within a distance of $d$, a particle diameter with a magnitude 
\begin{equation}
k_P \left(d - r_{ij}\right) + a_P \left(d - r_{ij}\right)^3 - \Gamma \hat{r}_{ij} \cdot \delta \vec{v}_{ij} \ \ .
\end{equation} 
These interactions are referred to as pair interactions and are censored between bonded particles.
This force consists of a harmonic repulsion with stiffness $k_P$, an analogous damping force with equal strength $\Gamma$, and an additional anharmonic repulsive force with strength $a_P$.
The anharmonic term is added to ensure the surfaces of grains remain impenetrable at very high packing fractions and that particles from separate grains do not overlap.
We equate the two linear stiffness terms $k \equiv k_B = k_P$ and define a unit of time $\tau \equiv \sqrt{m/k}$.
Simulations then use $\Gamma = \sqrt{k m}$, $a_P = 50 k/d^{2}$, and various values of $a_B$.

%==============================================================%
\subsection{Calibration}
\label{sec:calibrate}
%==============================================================%

To create an initial network of bonded particles representing a bulk solid, we randomly fill cubic boxes of linear size $L$ with enough monodisperse spheres of diameter $d$ to fill a volume fraction of $0.64$.
Periodic boundary conditions are used along all dimensions.
Overlaps between particles are first removed by applying an additional viscous damping forces and running overdamped dynamics.
All damping, both viscous and pairwise, is then removed and particles are thermalized to a temperature of $0.1$ by generating random velocities.
Simulations are run for a time of $1000 \tau$ before pairwise damping is restored and simulations are quenched over $3000 \tau$.
This protocol results in a disordered packing of particles near the jamming threshold.
Bonds are finally generated between all neighboring particles within a distance of $1.5 d$ resulting in each particle having $\sim 15$ bonds and the velocities of particles are zeroed. 
Trajectories are numerically integrated using the velocity-Verlet algorithm and a timestep of $0.1 \tau$.

After generating cubic systems of size $L = 50 d$, elastic moduli are calibrated by both compressing and shearing the system to a volumetric and shear strain of $0.5\%$ as conducted in a separate work \cite{Clemmer2023b}.
A stress tensor is calculated as the sum of the virial and kinetic contributions
\begin{equation}
\sigma_{\alpha \beta} = -\frac{1}{V}
      \left( \sum_\mathrm{particles} v_\alpha v_\beta -
	  \sum_\mathrm{interactions} F \frac{r_\alpha r_\beta}{|\vec{r}|}   \right)
\label{eq:stress_tensor}
\end{equation}
where $V$ is the volume of the entire system, $\vec{v}$ is a particle velocity, $\vec{r}$ is the displacement between two particles, and the summation across interactions includes bond and pair forces with magnitudes $F$. 
Since these systems are currently fully-bonded, there are no pair interactions between particles.
Initially there are no forces between bonds which are at the equilibrium reference length and $\sigma_{\alpha \beta} = 0$. 
As the system deforms, bonds exert forces and the pressure grows.
In compression the mean pressure $P \equiv -(\sigma_{xx} + \sigma_{yy} + \sigma_{zz})/3$ grows linearly with volumetric strain and in simple shear the shear stress grows linearly with shear strain.
Data is fit using a least means square linear regression to estimate the bulk and shear moduli. 

\begin{figure}
\begin{centering}
	\includegraphics[width=0.9\columnwidth]{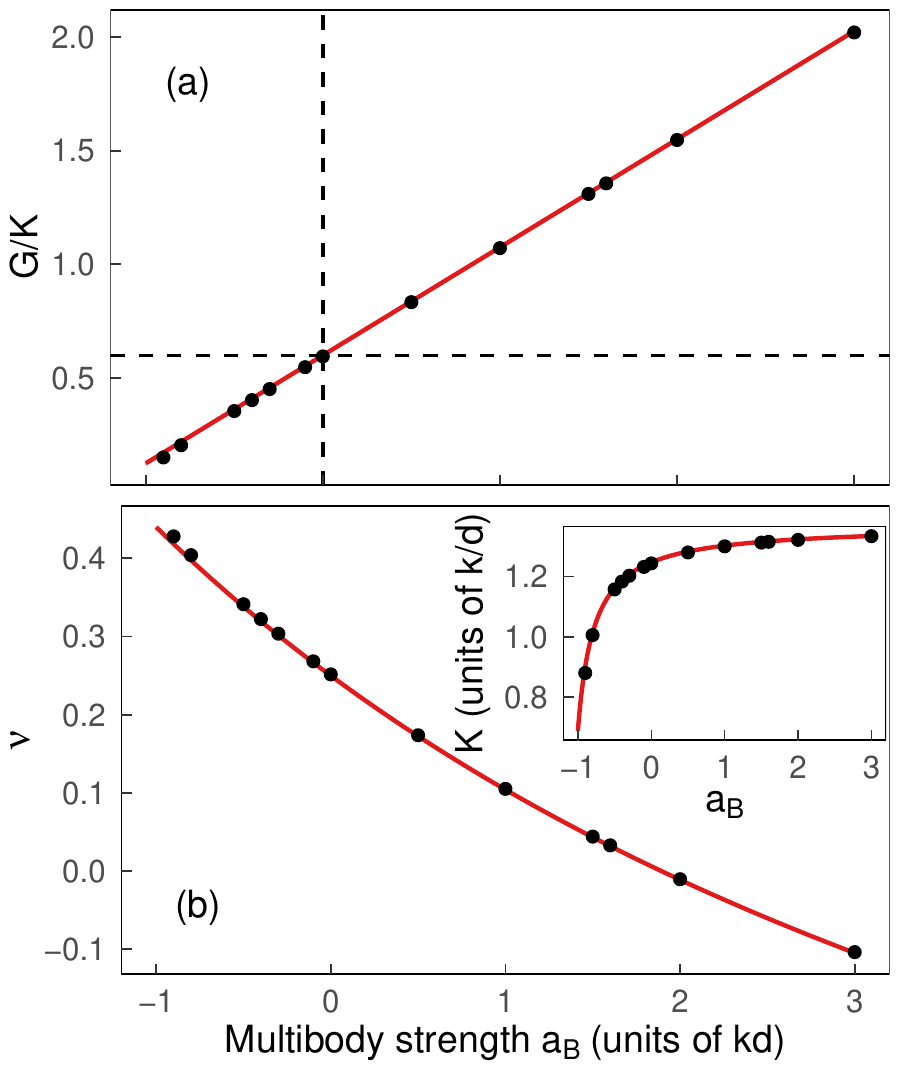}
	\caption{(a) The ratio of the measured shear modulus $G$ to the bulk modulus $K$ as a function of the strength of the multibody term $a_B$. Dashed lines indicate $G/K = 0.6$ and $a_B = 0.0$. (b) The Poisson's ratio $\nu$ and bulk modulus $K$ (inset) vs. $a_B$. Red lines indicate fits described in the text.}
	\label{fig:moduli}
\end{centering}
\end{figure}

First, we consider the ratio of the shear to bulk moduli $G/K$, which grows approximately in proportion to the multibody force strength $a_B$ as shown in Fig. \ref{fig:moduli}(a). 
At $a_B = 0$, one expects a Poisson's ratio of $\nu = 1/4$ which corresponds to $G/K = 3/5$.
With this constraint, we fit a linear regression to this data and find $G/K \approx 0.6 + 0.475 a_B (k d)^{-1}$.
Notably, this implies the incompressible limit, $G/K = 0$ or $\nu = 1/2$, is near $a_B=-k d$, the point where the multibody force is approximately equal in magnitude to the harmonic force.
Therefore, the multibody term can both increase $\nu$ above 1/4 for $a_B <0$ and decrease $\nu$ below 1/4 for $a_B > 0$ as demonstrated in Fig. \ref{fig:moduli}(b).
Practically, however, we find it is still challenging to reach the true incompressible limit of $\nu = 1/2$ as systems can become unstable.

Ideally, the bulk modulus $K$ would be independent of $a_B$ such that $a_B$ only controls the shear resistance.
However, we do find some dependence of $K$ on $a_B$ in the inset of Fig. \ref{fig:moduli}(b) as $K$ drops near the incompressible limit which is likely related to the previously mentioned instability. 
As further discussed in the appendix, particles under compression have greater nonaffine displacements in the incompressible limit suggesting they adjust their positions to minimize total compressive bond forces using the multibody term.
Practically, we find this effect can be largely avoided either by: a) using a regular lattice or b) adding a constraint that the net multibody force exerted by a particle on its neighbors is zero.
Since these options either introduce anisotropy or additional computational costs, we opt to keep the current formulation and recognize that this just entails a separate calibration for $K$.
Data was empirically fit using $K = 1.37 k/d - 0.15 k^2 (a_B + 1.22 k d)^{-1}$.
While this calibration is used to approximate moduli in the text, if there were a change in the initialization protocol or bond creation criteria then this procedure would need to be rerun.

To quantify the accuracy of this approach, we studied the convergence of both the calibrated parameters and deviations from linear elastic solutions as a function of increasing resolution.
In a BPM, increasing the resolution of a simulation is equivalent to increasing the number of particles used to represent a unit of volume. 
In finite systems, there is some variation in macroscopic moduli across different random realizations that decrease with increasing system size.
Testing cubic systems of different sizes $L^3$, we find that the measured bulk and shear moduli both converge to an infinite-system-size limit at a rate of $\sim L^{-2}$. 
Furthermore, variations in measured moduli between random realizations of the same system size decay as $L^{-3/2}$ as larger systems have less variation between samples, likely reflecting a typical $\sqrt{N}$ scaling where $N$ is the number of independent volume elements.
Lastly, local deviations in the strain of particles decay as $L^{-1}$ with increasing resolution. 
This simply implies the non-affine displacement of particles is constant and independent of system size.
However, many materials are disordered on some length scale leading to localization effects that are critical in their mechanical response suggesting this inherent feature of BPMs is not necessarily inaccurate.
For the resolutions considered in this work, grains may exhibit variations in macroscopic parameters of approximately $1\%$ and the local strains of particles in the grains may deviate by approximately $0.1\%$ to $0.01\%$ on average from the continuum elastic solution at small strains.
More details on these tests are provided in the appendix.

%==============================================================%
\subsection{Grain generation and compaction protocol}
\label{sec:simultions}
%==============================================================%

\begin{figure}
\begin{centering}
	\includegraphics[width=\columnwidth]{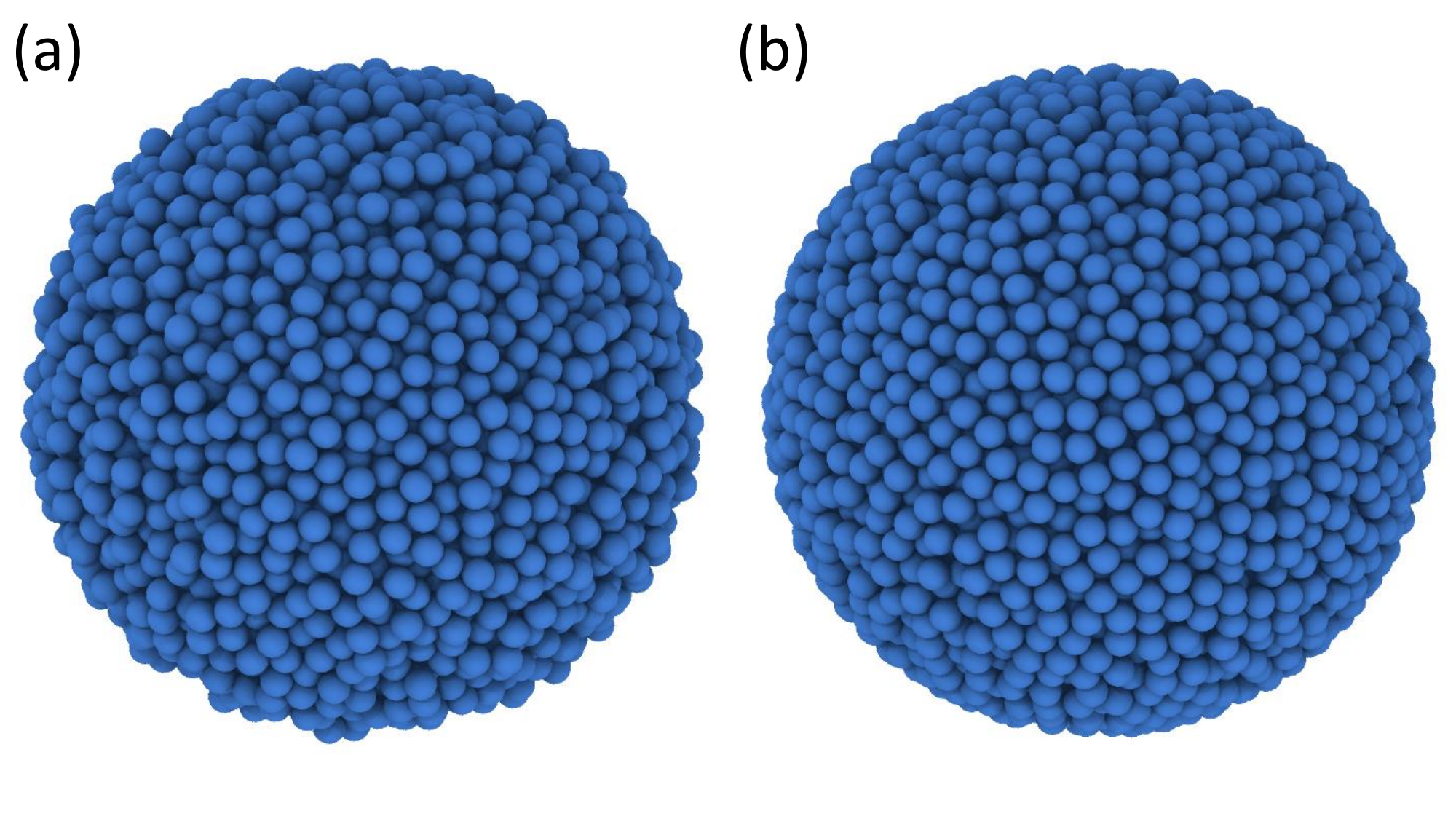}
	\caption{Rendered images of (a) rough and (b) smooth grains consisting of $\sim 5,000$ BPM particles each.}
	\label{fig:grain_geom}
\end{centering}	
\end{figure}

To create spherical grains of bonded particles, we used two protocols.
The first approach is to take the disordered packing of particles described above and cut out monodisperse spherical regions.
Any particles with a position beyond a distance of ten particle diameters from the center of the grain were deleted, resulting in $\sim$5,000 particles per grain.
As seen in Fig. \ref{fig:grain_geom}(a), this results in a fairly irregular surface and we therefore label this the rough model.
To create a smoother grain, before constructing bonds we first cut out a sphere of packed particles and construct a spherical repulsive wall around the packing with a harmonic potential with unit stiffness and unit cutoff that applies an inward force to any particles in contact. 
The wall is initially at a distance large enough to avoid any contacts but its diameter then oscillates with an exponentially decaying amplitude to reorder particles on the surface while simultaneously applying viscous forces to overdamp dynamics before constructing bonds. 
This process essentially vibrates grains into smoother shapes and is referred to as the smooth model (Fig. \ref{fig:grain_geom}[b]).
As friction only emerges in a point-particle-based BPM through the geometric corrugation on the surface of grains, these two geometries represent a high and low frictional grain, respectively.

\begin{figure}
\begin{centering}
	\includegraphics[width=0.9\columnwidth]{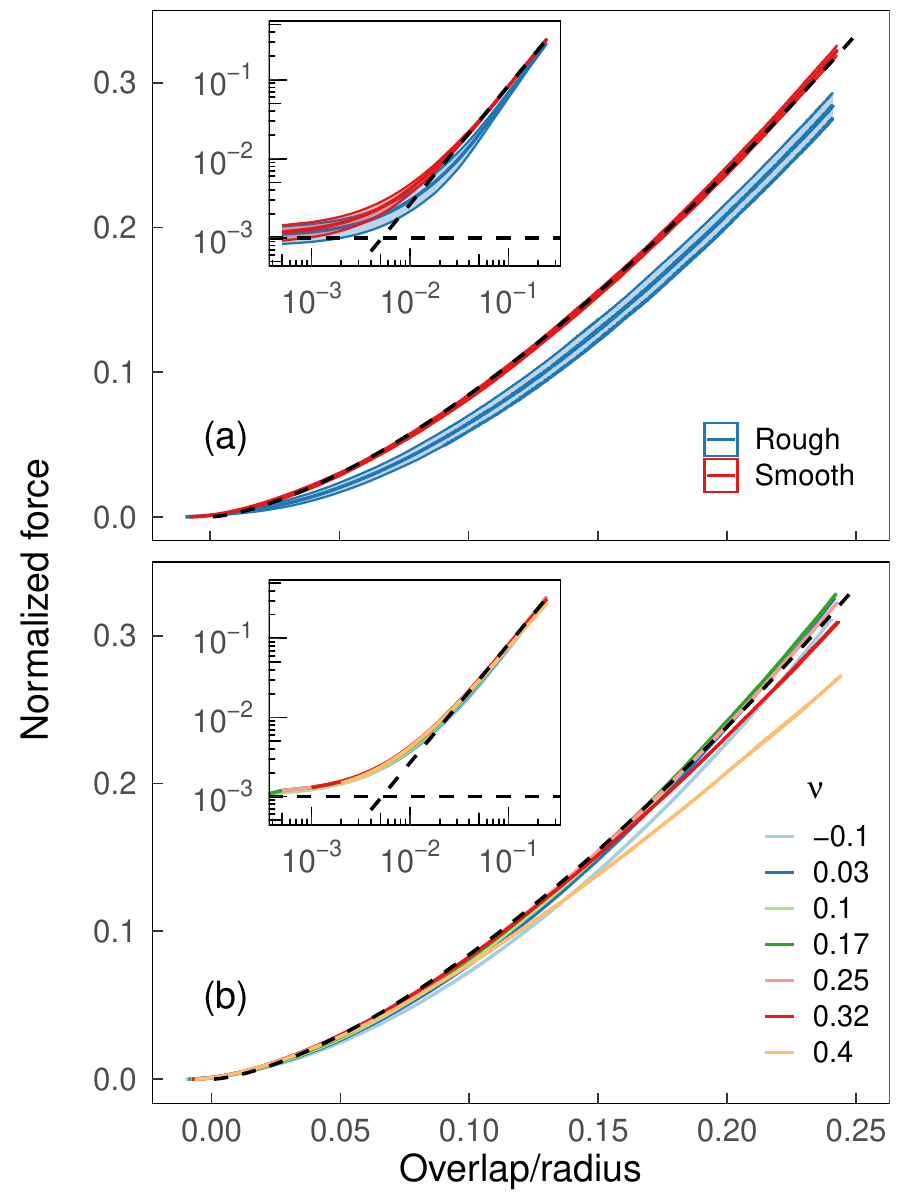}
	\caption{(a) Average contact forces normalized by $E_\mathrm{eff} R^2$ as a function of the normalized overlap for rough grains (blue) and smooth grains (red) in comparison to analytic Hertzian contact force (dashed, black line) at a Poisson's ratio of $\nu = 1/4$ ($a_B = 0$). Ribbons around the average force represent one standard deviation. (b) Contact forces for smooth grains at the indicated $\nu$. Log-log plots of both panels are provided as insets with dashed lines representing the force cutoff of $10^{-3} E_\mathrm{eff} R^2$ and a $3/2$ power law.}
	\label{fig:hertz}
\end{centering}
\end{figure}

To compare normal contact forces between BPM grains to Hertzian contact theory, computed force-displacement relationships are obtained by compressing BPM spheres between symmetrically displaced, mathematically smooth walls.
Walls are displaced at a constant rate of $5 \times 10^{-4} \tau^{-1}$. 
BPM particles interact with the walls using a harmonic potential with unit stiffness and unit cutoff distance to prevent interpenetration. 
The reported force is the average of the total normal force on each wall mediated by the wall-particle interaction. 
The centers of mass of the spheres are tethered to their initial positions with a soft spring to prevent unbounded lateral translation, which originates from the lack of friction between walls and spheres. 
The effect of this lateral constraint on the normal force was negligible. 
The ensemble average for rough spheres was calculated using 292 distinct BPM spheres; for smooth spheres, the ensemble average was calculated using 50 randomly rotated sphere configurations to alter the wall-sphere contact points.
Normalized forces for various grain geometries and material properties are plotted as a function of a radius-normalized overlap $\delta/R$ in Fig. \ref{fig:hertz} based on the expected Hertz force,
\begin{equation}
\frac{3}{8 R^2 E_\mathrm{eff}} f(\delta) = \left(\frac{\delta}{R}\right)^{3/2}
\label{eq:hertz}
\end{equation}
where $R$ is the grain radius, $E_\mathrm{eff} \equiv 3 K (1 - 2 \nu) /(2 - 2\nu^2)$ is an effective Young's modulus, and the overlap $\delta$ is defined relative to the first instance the force exceeds a magnitude of $10^{-3} E_\mathrm{eff} R^2$.
Due to the rough surface, there is not an obvious definition of $R$ and we therefore simply assume $R = 10 d$ for both smooth and rough grains in the following analysis.

Comparing forces for rough and smooth grains at $\nu = 1/4$ (Fig.~\ref{fig:hertz}[a]), we see markedly strong agreement for smooth grains with minimal variation while forces for rough grains are notably weaker with more variation between grains and contact location.
This is to be expected as the roughness of the surface implies contact depends on the exact location of the outermost surface particles, reminiscent of asperities on rough surfaces~\cite{Luan2005, Luan2006, Pastewka2016} which also cause deviations from Hertz theory.
Stronger agreement might be found if one tailored the force threshold to define the initial contact for each system or accounted for the roughness in estimating a radius.
Focusing on smooth grains, we also see good agreement with the analytic expectation across a wide range of Poisson's ratios serving as an important verification of the multibody term in a system with free surfaces and large deformations (Fig.~\ref{fig:hertz}[b]).
Deviations are slightly larger at the extremes of $\nu$ but are much smaller than the change in the magnitude of un-normalized forces.

\begin{figure*}
\begin{centering}
	\includegraphics[width=\textwidth]{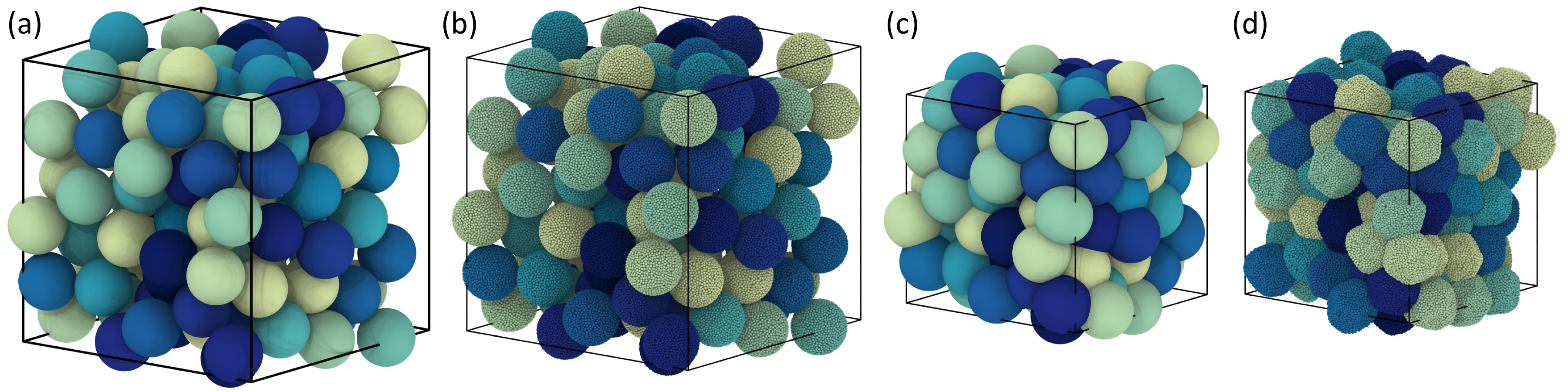}
	\caption{(a) Loose packing of $N = 128$ DEM grains, or one eighth of the actual system size. (b) The corresponding packing of BPM grains. The same (c) DEM and (d) BPM systems after compacting to a volumetric strain of $\sim 60\%$.}
	\label{fig:system_gen}
\end{centering}	
\end{figure*}

The next step is to generate packings of grains and compress them.
To reduce computational costs, templates of loose packings of 1,024 grains with $57\%$ volume fraction are generated using traditional DEM simulations  \cite{Silbert2001}.
An example with 128 grains is rendered in Fig. \ref{fig:grain_geom}(a).
These templates are then used to locate BPM grains, mapping particles and bonds across periodic boundaries as necessary.
Systems are relaxed and slightly expanded to break all contacts before resetting the equilibrium lengths of bonds to create stress-free granular states just below jamming as seen in Fig. \ref{fig:system_gen}(b).

Systems are then systematically compacted using a protocol aimed at sampling a wide range of logarithmically distributed pressures.
First, systems are jammed at a pressure of $P_T = 10^{-6}$ by running dynamics for a total time $t$ of $2 \times 10^5$ using a linear pressure control with a gain constant of $0.1 d/k$ that isotropically expands or contracts the box. 
The box is then istropically compressed at a constant true strain rate $\dot{\epsilon}$ along each dimension.
As box lengths evolve, particle positions are shifted to track their relative position in the simulation box.
During the constant strain rate compaction, the strain rate is initially $\dot{\epsilon} = 10^{-9} \tau^{-1}$, however, every $2 \times 10^5$ units of time it is gradually incremented
until it reaches $10^{-6} \tau^{-1}$ to probe a wide range of states near and far from the jamming limit.
At all rates, the average kinetic energy per particle remains small, always below $10^{-6} k d^2$, and there are minimal inertial effects.
A selection of systems were run using a purely pressure-controlled protocol at a much slower rate and no identifiable differences were found in results.
A sample compressed system is rendered in Fig. \ref{fig:system_gen}(c).
Five random realizations were run to average results.

%==============================================================%
\subsection{Equivalent discrete element method simulations}
\label{sec:dem}
%==============================================================%

As a comparison, we also simulated the compaction of various traditional DEM contact models.
Normal contact forces are Hertzian with damping forces based on the formulation by Tsuji, Tanaka, and Ishida \cite{Tsuji1992}. 
Tangential forces are calculated using a Mindlin no-slip solution \cite{Mindlin1949} with a sliding friction coefficient $\mu_s$.
Rolling and twisting friction are applied using spring-dashpot-slider models by Luding \cite{Luding2008} and Marshall \cite{Marshall2009} with rolling and twisting friction coefficients $\mu_r$ and $\mu_t$, respectively.

To match DEM to BPM simulations, we varied three parameters: $\mu_s$, $\mu_r$, and $\mu_t$. 
Instead of performing mechanical tests on individual BPM grains to measure frictional forces at contacts, we opted to simply adjust the DEM friction coefficients to achieve similar jamming packing fractions $\phi_c$ \cite{Silbert2010, Santos2020}.
We find four sets of coefficients $\{\mu_s, \mu_r, \mu_t\}$ that approximately reproduce $\phi_c$ for both the rough and smooth BPM grains, $\{0.3, 0, 0\}$ and $\{0.15, 0.1, 0.1\}$ for rough grains (higher friction) and $\{0.15, 0, 0\}$ and $\{0.1, 0.05, 0.05\}$ for smooth grains (lower friction).
This provides a DEM equivalent with and without rolling and twisting friction to compare to each type of BPM grain.

%==============================================================%
\subsection{Implementation in LAMMPS}
\label{sec:lammps}
%==============================================================%

To model a large number of high resolution grains, we implemented a package in the particle dynamics codebase LAMMPS for modeling BPM systems \cite{Thompson2021}.
The package is available open source with the main LAMMPS distribution and is designed to support a wide range of bond styles and, due to the design of LAMMPS, is easy to modify and expand.
Currently, the package includes an implementation of a point-particle-based model as used in this work as well as a more common model using DEM particles with rotational degrees of freedom and bonds that transmit shear forces and torques \cite{Wang2009}. 
Features of the package include generalized methods for bonds to store data (such as a reference state or strain history to model plasticity \cite{Clemmer2023}), communicate with neighbors to calculate multibody interactions, break under various loading conditions (to model fracture or fragmentation \cite{Clemmer2022}), and optionally either overlay or censor pair forces between bonded particles.
All capabilities were developed with parallel efficiency in mind. 
DEM simulations were performed using the GRANULAR package.

\section{Results and discussion}
\label{sec:results}

%==============================================================%
\subsection{Deviations from jamming and friction dependence}
\label{sec:standard}
%==============================================================%

We first set out to understand how and when results deviate from the low-pressure scaling behavior near jamming and measure friction dependence.
Here, we narrow attention onto BPM systems without the multibody term, setting $a_B = 0.0$ or $\nu = 1/4$, and only consider standard metrics of the jamming transition: pressure $P$, packing fraction $\phi$, and the average coordination number $Z$.
This section also includes a comparison between BPM and DEM simulations to quantify errors related to omitting the internal elasticity of grains.

In DEM simulations, $\phi$ is typically defined as:
\begin{equation}
\phi_\mathrm{sphere} = \frac{1}{V} \sum_\mathrm{grains} \frac{4}{3} \pi R_i^3 
\label{eq:phi_naive}
\end{equation}
where $V$ is the volume of the entire system and $R_i$ is the radius of grain $i$. 
However, this definition is inadequate for BPM simulations as grains inherently have surface roughness, and therefore are not perfectly spherical, and can deform. 
Therefore, we alternatively calculate the volume fraction using Monte Carlo integration, $\phi_\mathrm{MC}$. 
Within the simulation cell, random points are uniformly sampled and tested for collisions with all particles in the simulation (where each particle has a radius of $d/2$) and with the convex hull of each grain to include the internal volume of gaps between particles.
Using $10^7$ random points, uncertainty in $\phi_\mathrm{MC}$ is limited to the fourth digit. 

\begin{figure}
\begin{centering}
	\includegraphics[width=0.9\columnwidth]{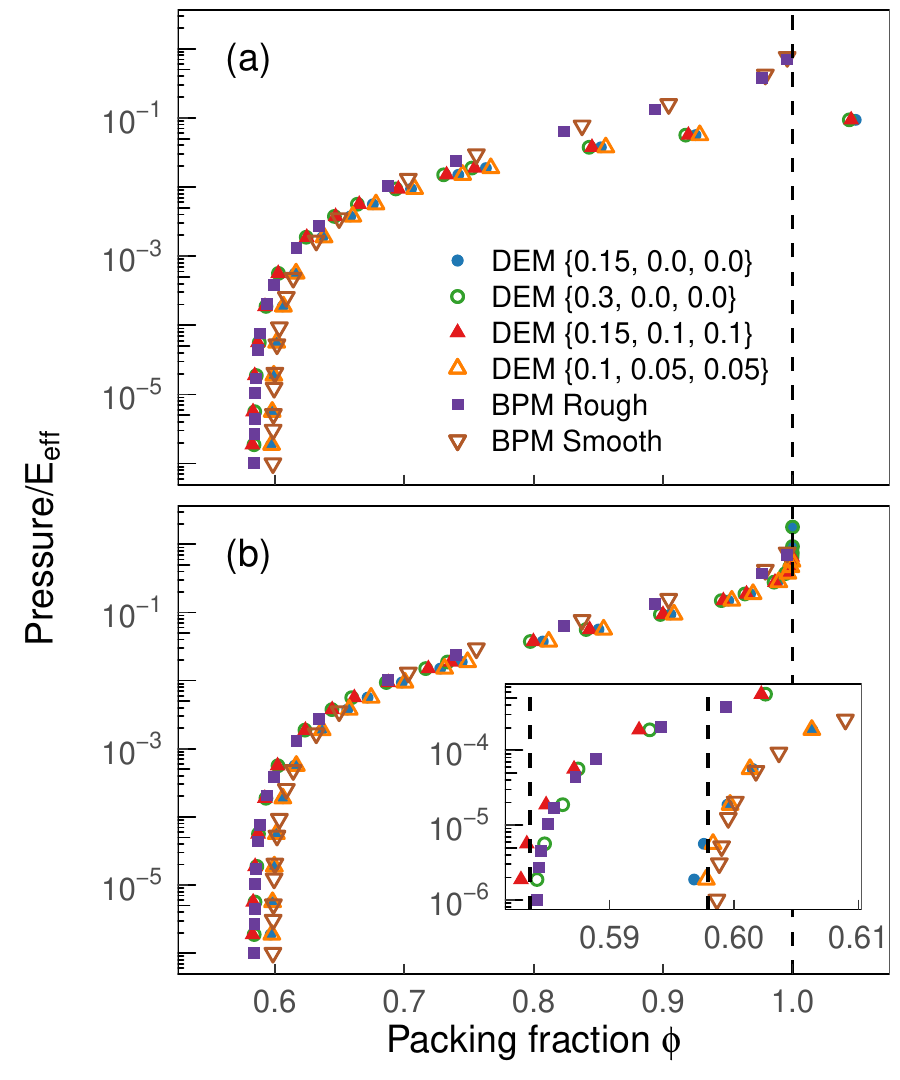}
	\caption{Pressure normalized by the effective modulus as a function of packing fraction for the indicated DEM or BPM systems. For DEM systems, friction coefficients are reported as $\{\mu_s, \mu_r, \mu_t\}$ and the packing fraction is defined as (a) $\phi_\mathrm{sphere}$ and (b) $\phi_\mathrm{MC}$. For BPM systems, both panels report $\phi_\mathrm{MC}$. The inset in panel (b) highlights curves near jamming. Dashed lines indicate $\phi = 0.5836, 0.5979$, and $1.0$.}
	\label{fig:PvPhi}
\end{centering}
\end{figure}

Initially, both DEM and BPM systems start just below the jamming transition at a small but non-zero pressure. 
To simplify comparisons, reported pressures are normalized by the effective modulus for a Hertzian contact $E_\mathrm{eff}$, Eq.~\eqref{eq:hertz}.
Using the above definitions, we find $\phi_\mathrm{MC} \sim 0.60$ for smooth particles and $\phi_\mathrm{MC} \sim 0.58$ for rough particles at pressures near $P/E_\mathrm{eff} = 10^{-6}$.
Very similar values of $\phi_\mathrm{sphere}$ are found in the four sets of DEM packings as seen in compaction curves in Fig. \ref{fig:PvPhi}(a).
Differences between DEM and BPM values of $\phi$ are initially no greater than $\pm 0.003$ for matched systems.
As systems compact, the pressure rises and there are distinct high- and low-friction curves.
This continues until a packing fraction of $\sim0.7$ where compaction curves for the two BPM systems coalesce as friction becomes irrelevant.
This is expected at large packing fractions since tangential forces become less relevant as particles become fully confined by large normal forces and can no longer rotate.
Meanwhile, results for DEM and BPM simulations diverge with each method following a separate compaction curve.
Similar deviations at high pressures have been identified between DEM and FEM-based simulations \cite{Cardenas-Barrantes2022}.

At very high pressures in DEM simulations, deficiencies in the above definition of $\phi_\mathrm{sphere}$ are apparent in Fig. \ref{fig:PvPhi}(a) as it exceeds unity.
As noted in other works on deformable particles \cite{Boromand2018, Boromand2019}, this definition does not account for overlapping regions of DEM particles such as those seen in Fig. \ref{fig:system_gen}(c) and overestimates $\phi$.
While this effect is negligible in the typical hard-particle limit studied using DEM, it is significant in the high pressure limit.
Therefore, $\phi_\mathrm{MC}$ was additionally calculated for DEM systems using $10^6$ randomly sampled locations to avoid this issue.
As seen in Fig. \ref{fig:PvPhi}(b), at low pressures results are virtually independent of the definition of $\phi$ but the expected upturn in pressure at the maximum packing fraction of 1.0 is now seen in DEM results.
Comparing DEM and BPM results, the shape of compaction curves is now qualitatively quite similar and we see errors associated with DEM are limited to within about a factor of two.
Given the fact that the grain-grain normal forces are well described by the Hertzian model up to significant overlaps of 25\% of the radius (Fig. \ref{fig:hertz}), this may not be too surprising.
Of course, DEM simulations fail to account for elastic interactions between contacts such that more advanced nonlocal DEM contact models \cite{Gonzalez2012, Giannis2021} might provide even stronger agreement.
Additionally, both the DEM and BPM simulations assume linear elasticity while many real materials would exhibit non-linear effects at these large strains such that results may further diverge.
Future work testing the compaction of a non-linear elastic granular material, such as a neo-Hookean model, using BPM simulations would prove useful.

\begin{figure}
\begin{centering}
	\includegraphics[width=0.9\columnwidth]{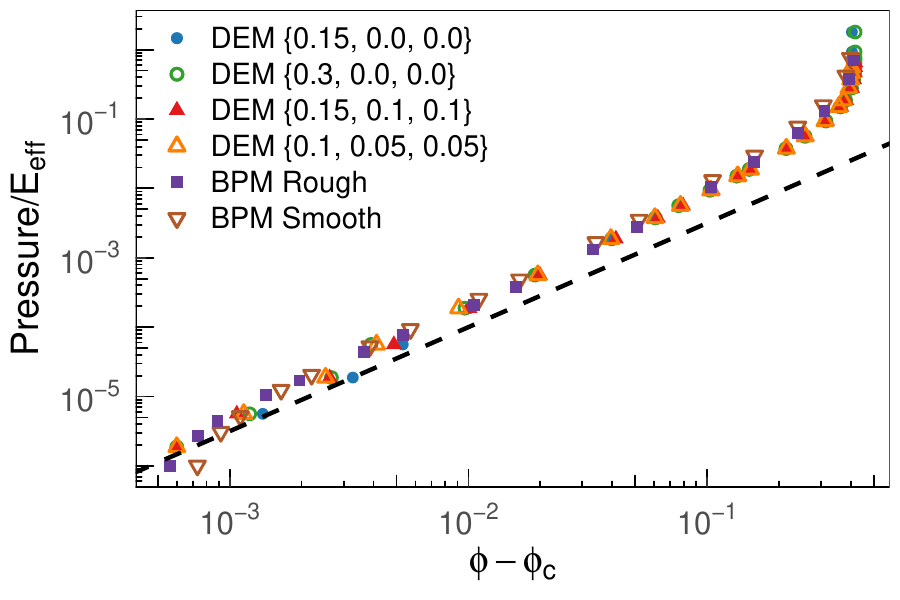}
	\caption{Normalized pressure as a function of $\phi - \phi_c$ for the indicated systems with $\phi = \phi_\mathrm{MC}$. The dashed line has a slope of $3/2$.}
	\label{fig:PvdPhi}
\end{centering}
\end{figure}

To compare compaction curves to the expected power-law scaling near the jamming point, we first estimate the jamming packing fraction $\phi_c$. 
For the remainder of the article, the volume fraction is exclusively calculated using Monte Carlo integration and the $\mathrm{MC}$ subscript is dropped on $\phi$.
Near the jamming transition, the pressure grows as $(\phi-\phi_c)^{3/2}$ for Hertzian contact forces \cite{OHern2003}.
For each system, $\phi_c$ is estimated by calculating the minimum measured packing fraction and adjusting it to maximize this power-law scaling at small $\phi-\phi_c$ in Fig. \ref{fig:PvdPhi}.
For the BPM systems, we estimate $\phi_c = 0.5979$ and $0.5836$ for the low- and high-friction cases, respectively.
This scaling persists to values of $\phi-\phi_c \sim 10^{-1}$ at which point the growth in pressure accelerates as $\phi$ approaches unity.
This threshold also approximately corresponds to the point where friction becomes irrelevant and BPM and DEM results deviate.

To confirm BPM results are converged with the resolution of grains, simulations were run for smooth grains with radii of $2.5$, $5$, and $7.5d$. 
At small resolutions, compaction curves are shifted to smaller packing fractions but approach the above results with increasing resolution.
Differences in results between systems with grain radii of $7.5d$ and $10d$ are less than the size of a symbol in Fig. \ref{fig:PvPhi} implying there are minimal finite resolution effects.
Further tests demonstrating the convergence of BPM simulations with increasing resolution are presented in the appendix.

\begin{figure}
\begin{centering}
	\includegraphics[width=0.9\columnwidth]{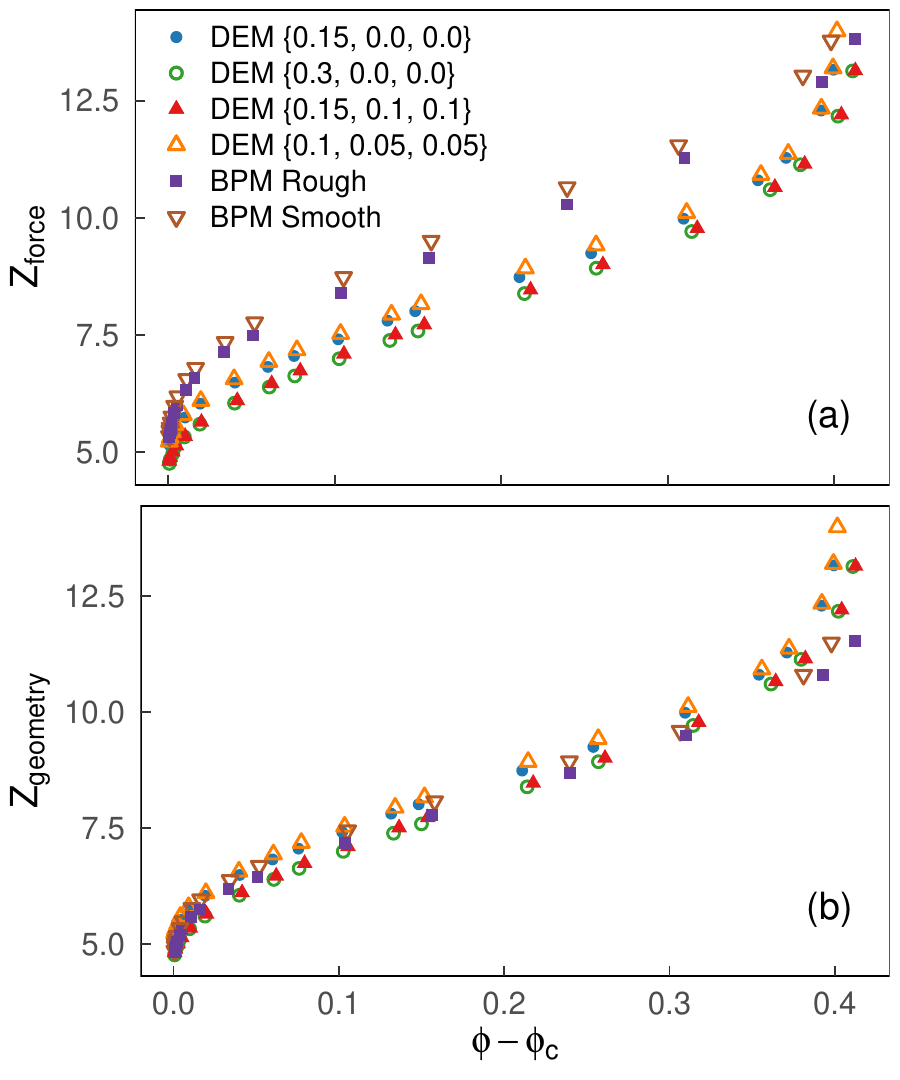}
	\caption{Average coordination number as a function of the distance to jamming for the indicated systems. For BPM systems, contacts between grains are identified by (a) identifying contacts between particles $Z_\mathrm{force}$ and (b) using a naive geometry $Z_\mathrm{geometry}$. Data at $Z > 14$ is truncated to improve visibility. }
	\label{fig:ZvPhi}
\end{centering}
\end{figure}

In addition to the pressure, another key metric of the jamming transition is the coordination number. 
In this work, we calculate an average coordination number $Z$ after excluding rattlers where rattlers are defined as undercoordinated grains with two or fewer contacts.
While $Z$ is trivially measured in DEM systems by counting contacts, in BPM systems there is not an obviously correct metric.
One approach is to define a contact between any two grains which have constituent particles in contact which we label $Z_\mathrm{force}$.
Using this definition in Fig. \ref{fig:ZvPhi}(a), we see higher coordination numbers in systems with reduced friction, as expected.
More importantly, we also see a systematically higher coordination number in both BPM systems relative to DEM systems near jamming which could be due to the surface roughness of BPM grains.
Furthermore, $Z_\mathrm{force}$ may be larger due to the fact that two particles can be in contact but not exert a stabilizing force on the grains.
For instance, one could imagine a contact could consist of four particles, two on each grain, which exert equal and opposite tangential forces.

In analogy to DEM, one could alternatively use a geometric approximation to identify contacts in a BPM system by detecting when the center of mass of two grains is within a fixed distance.
To choose this distance, we calculate the maximum distance between the center of mass of a grain and any particle in the grain and then averaging across all grains.
Doubling this and adding the interaction distance between particles, $d$, we find a cutoff distance of $20.27 d$ for smooth grains and $20.53 d$ for rough grains. 
This approximates the diameter of a sphere that would encase a grain.
Using this cutoff, the average coordination number $Z_\mathrm{geometry}$ is quite similar to DEM systems as seen in Fig. \ref{fig:ZvPhi}(b). 
The curves for $Z_\mathrm{geometry}$ for BPM systems lie between the high and low friction DEM systems near the jamming transition but do not grow as rapidly as $\phi \rightarrow 1.0$.
Of course, this definition does not account for the elastic deformation of grains.

\begin{figure}
\begin{centering}
	\includegraphics[width=0.9\columnwidth]{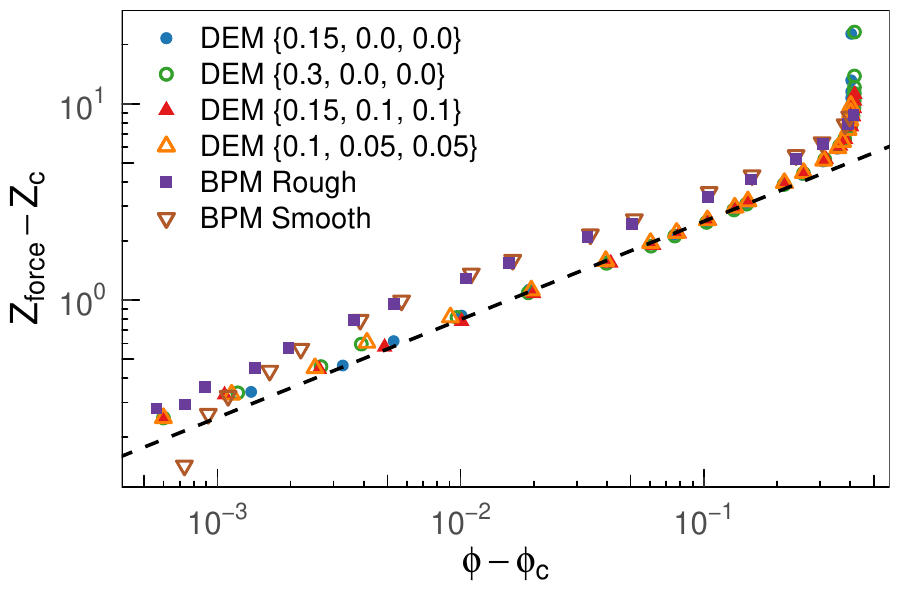}
	\caption{The data in Fig. \ref{fig:ZvPhi}(a) is plotted after subtracting the limiting coordination number at jamming. The dashed line has a slope of $1/2$.}
	\label{fig:ZvdPhi}
\end{centering}
\end{figure}

Regardless of the definition of the coordination number, near the jamming transition we find the expected scaling behavior.
Jamming corresponds to an isostatic point where the number of constraints on the system is equal to the number of degrees of freedom.
For frictionless systems, this implies jamming occurs at an average coordination number $Z$ of $Z_c = 6$ in 3D. 
When friction is introduced, $Z_c$ decreases as fewer contacts are needed to stabilize the packing \cite{Silbert2010}, however, the excess coordination number $Z - Z_c$ is always found to grow as a $(\phi - \phi_c)^{1/2}$ \cite{OHern2003, Silbert2010}.
This is found to be true regardless of frictional strength \cite{VanHecke2010, Silbert2010} and has been found to persist to relatively high pressures in experiments \cite{Bares2023}.
Here, we measure $Z_c$ by calculating the smallest value of $Z$ and adjusting it to maximize the power-law domain at low packing fractions finding $Z_c \sim 5.05$ for rough grains and $Z_c \sim 5.2$ for smooth grains. 
In comparison for DEM systems, we measure $Z_c \in [4.52, 4.57]$ and $Z_c \in [4.91, 4.98]$ for high- and low-friction calibrations, respectively.
In Fig. \ref{fig:ZvdPhi}, the expected scaling is seen in both DEM and BPM systems where the latter uses the $Z_\mathrm{force}$ definition. 

\begin{figure*}
\begin{centering}
	\includegraphics[width=0.95\textwidth]{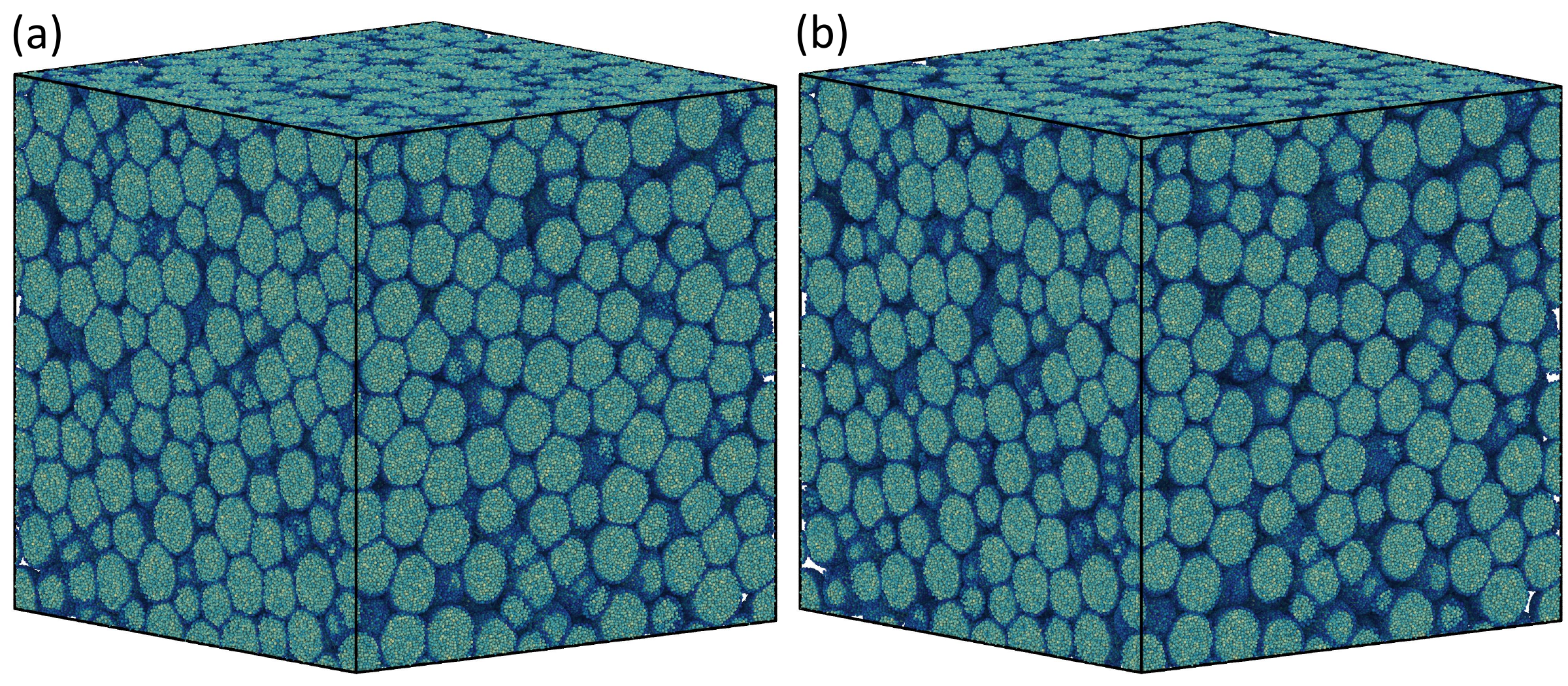}
	\caption{Images of full-sized BPM simulations of 1,024 smooth grains with values of (a) $a_B = -0.8 k d$ or $\nu = 0.40$ and (b) $a_B = 3.0 k d$ or $\nu = -0.11$ compacted to an equivalent volume with packing fractions of $\phi = 0.86$ and $0.80$, respectively. Particles are colored by their number of bonds to highlight grain surfaces and remapped across periodic boundaries to highlight cross sections.}
	\label{fig:example_compaction}
\end{centering}
\end{figure*}

Under compaction, we find that BPM simulations reproduce the expected scaling near the jamming transition until deviations set in at $\phi - \phi_c \sim 0.1$. 
This transition also corresponds to a limit where friction becomes less important and the internal deformation of grains becomes relevant as compaction curves at different friction coefficients coalesce and DEM and BPM results diverge. 
However, if one considers analogous metrics, the predictions from DEM are still within a factor of two from the BPM simulations suggesting suggesting results from DEM simulations can still be somewhat informative far from the hard-sphere limit assuming systems are linear elastic.
However, as we explore in the following section, there is certain information that DEM simulations cannot provide.

%==============================================================%
\subsection{Impact of Poisson's ratio on grain deformation}
\label{sec:alternative}
%==============================================================%

Next we seek to investigate the role of elasticity in the high pressure compaction of grains. 
In traditional DEM simulations, Poisson's ratio only affects the stiffness of Hertzian contact force. 
Thus, DEM simulations produce trivially identical results as stresses are simply scaled by a constant factor, $E_\mathrm{eff}$. 
Therefore, in this section we focus only on BPM results and vary $a_B$ to control Poisson's ratio $\nu$ in the low friction limit using the smooth grain construction.

As a visual example, a highly compacted system with $\nu = 0.40$ in panel (a) and $\nu = -0.11$ in panel (b), are rendered at equal volumes in Fig. \ref{fig:example_compaction}.
The shape of grains visibly differ.
At a high Poisson's ratio near the incompressible limit, grains have significantly deformed from their original spherical shape, flattening at contacts. 
In contrast at a low Poisson's ratio in the auxetic limit, grains are more spherical with minimal change in shape.
This qualitative behavior is expected as smaller $\nu$ or large shear moduli increase the resistance to shape distortions. 
Since the two systems are rendered at equal volumes, this also implies the auxetic grains had to compress more to compensate for the lack of distortion.
Although it may be hard to identify by eye, the auxetic grains in panel (b) are indeed smaller than the nearly incompressible grains in panel (a) which have a minimal change in volume. 

\begin{figure}
\begin{centering}
	\includegraphics[width=0.9\columnwidth]{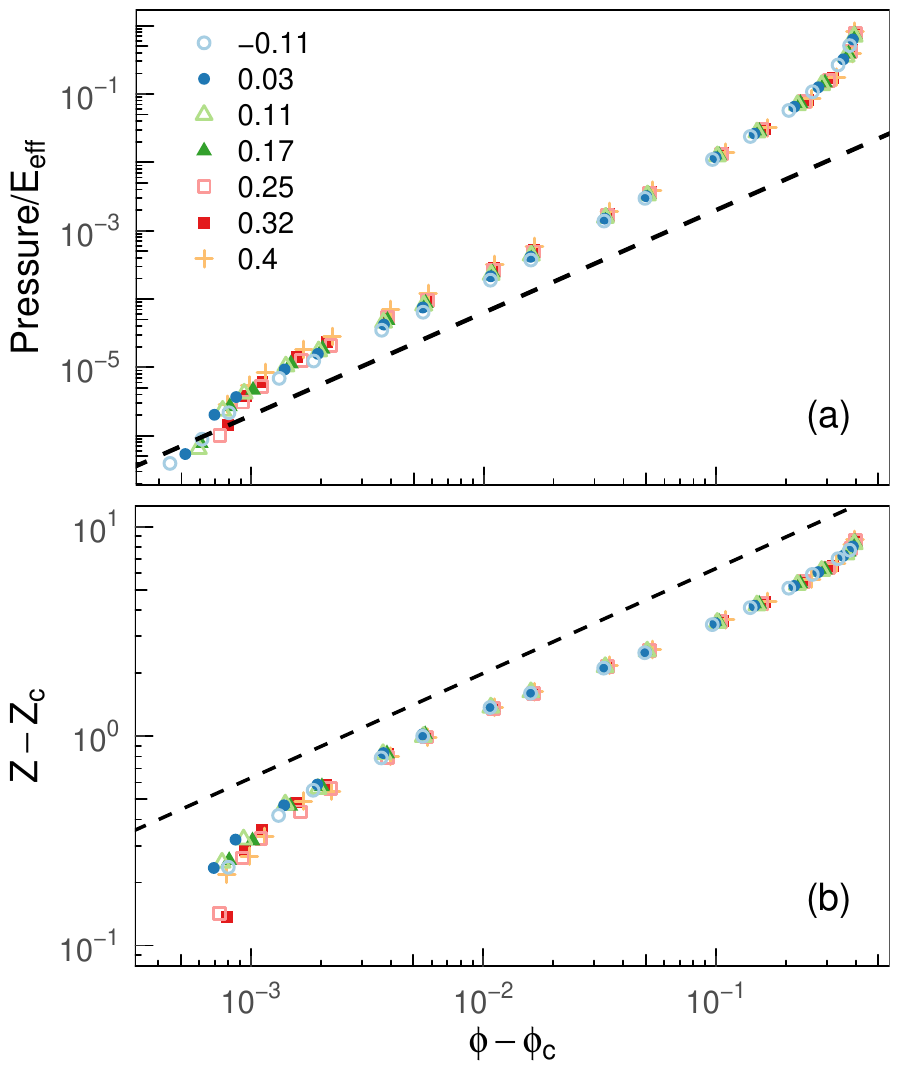}
	\caption{(a) The pressure as a function of the distance to the critical point for BPM simulations of smooth grains with the indicated $\nu$. (b) Excess coordination $Z_\mathrm{force}$ from jamming. Dashed lines have slopes of (a) 3/2 and (b) 1/2.}
	\label{fig:PZ_phi_nu}
\end{centering}
\end{figure}

Despite these observations, standard metrics from jamming exhibit minimal additional dependence on $\nu$ as seen in Fig. \ref{fig:PZ_phi_nu} which includes results for systems with smooth grains and $\nu$ between -0.11 to 0.4.
As in the above section, $\phi_c$ and $Z_c$ are estimated by roughly maximizing the power-law domain.
Across values of $\nu$, $\phi_c$ varies between 0.5977 to 0.5983, generally increasing with decreasing $\nu$.
$Z_c$ was fixed at 5.2 where $Z$ was defined as $Z_\mathrm{force}$. 
Aside from the standard scaling of the effective stiffness of the Hertz contact, results are largely independent of $\nu$ even to very high pressures with packing fractions close to unity.
In two dimensions, scaling by the effective stiffness has similarly been found by \citet{Vu2021} to reasonably describe differences in pressure at high packing fractions for systems with different Poisson's ratios.

As suggested in Fig. \ref{fig:example_compaction}, however, we do find quantitative differences in the deformation of grains: a fundamentally unresolved metric in DEM simulations.
For each grain, the convex hull is evaluated using the positions of all particles in the grain and is used to calculate both a volume $V_g$ and surface area $A_g$. 
Using these definitions, we considered two metrics: the volumetric strain 
\begin{equation}
\epsilon_{V_g} \equiv \frac{V_{0,g} - V_g}{V_{0,g}}
\end{equation}
where $V_{0,g}$ is the initial volume of the grain and the asphericity 
\begin{equation}
\alpha \equiv \frac{A_g^{3/2}}{6 \sqrt{\pi} V_g}
\end{equation}
where $\alpha = 1$ for a sphere and $\alpha > 1$ for an aspherical object, as used in other works on deformable grains.\cite{Boromand2018, Cardenas-Barrantes2022}

\begin{figure}
\begin{centering}	
\includegraphics[width=0.9\columnwidth]{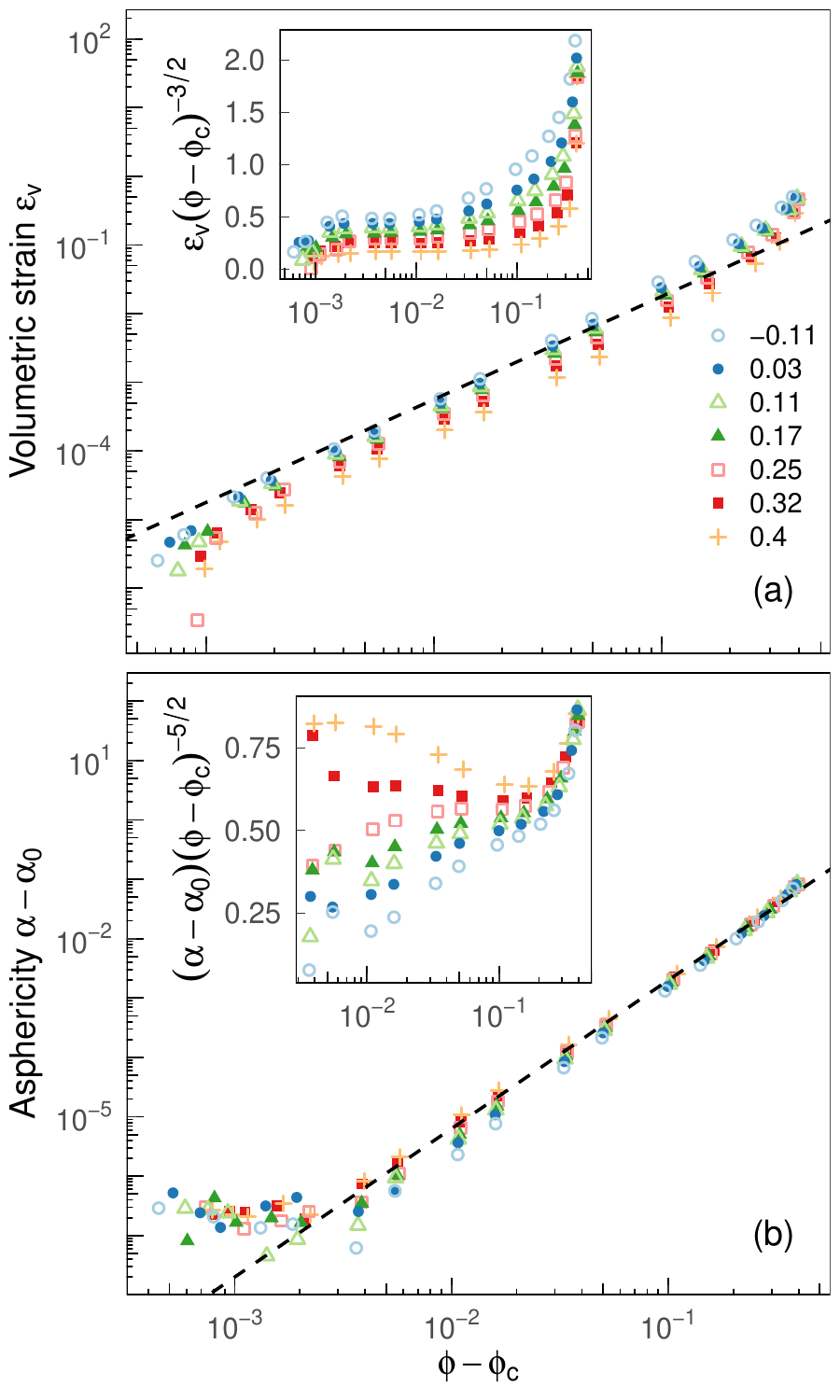}
	\caption{(a) Average volumetric strain per grain as a function of distance from jamming for smooth BPM systems at the indicated $\nu$. (b) Average asphericity for the same systems. Dashed lines have slopes of (a) 3/2 and (b) 5/2.}
	\label{fig:grain_deformation}
\end{centering}
\end{figure}

With increasing $\nu$, one expects less volumetric compression as one approaches the incompressible limit.
In Fig. \ref{fig:grain_deformation}(a), the average volumetric strain of a grain is found to grow as a power of the change in packing fraction with an exponent of $3/2$.
This implies the average volumetric strain of grains proportionally tracks the macroscopic pressure of the system which also grows as $(\phi-\phi_c)^{3/2}$ as seen in Fig. \ref{fig:PZ_phi_nu}(a). 
This scaling persists until $\phi - \phi_c \sim 0.1$ at which point the growth in $\epsilon_V$ accelerates as void spaces fill and $\phi \rightarrow 1$.
As $\nu$ increases, curves shift downward as there is a greater resistance to a change in volume. 
Scaling out the measured power law in the inset of Fig. \ref{fig:grain_deformation}(a), grains with $\nu = 0.4$ can experience approximately a quarter of the volumetric compression as grains with a near-zero Poisson's ratio at very large pressures. 

Alternatively as $\nu$ decreases, one expects stronger preservation of the shape of grains.
For instance in the auxetic limit, compression along one axis will induce compression along other axes minimizing distortion in the shape of an object.
Due to imperfections in representing a spherical object using a collection of point particles, at zero pressure there is some initial asphericity as $\alpha \sim 1.002$ in uncompressed grains.
Subtracting this initial value $\alpha_0$, the average change in asphericity is plotted in in Fig. \ref{fig:grain_deformation}(b). 
At low $\phi$, no significant change in asphericity is detected until $\phi-\phi_c \sim 10^{-3}$ above which $\alpha - \alpha_0$ grows as a power of $\phi-\phi_c$ with an exponent of about $5/2$ at $\nu = 1/4$.
This scaling also persists until $\phi-\phi_c \sim 0.1$ at which the growth in asphericity accelerates. 
This mirrors findings from \citet{Cardenas-Barrantes2022} which used the finite-element based NSCD and found a similar scaling behavior despite the completely distinct methodology. 

To accentuate deviations from scaling, the estimated power law is divided out in the inset of Fig. \ref{fig:grain_deformation}(b) revealing a splay in curves at intermediate $\phi$ emerges across values of $\nu$.
Grains in the auxetic limit experience less change in shape compared to grains near the incompressible limit, reflecting qualitative observations in Fig. \ref{fig:example_compaction}.
At $\phi -\phi_c \sim 10^{-3}$, $\alpha$ can vary by a factor of nearly 5.   
However, this difference narrows as data converges to a limiting response as $\phi \rightarrow 1$.
As spheres can never fully fill a volume, grains must distort in this limit.
To accommodate these two limits, data could be explained by a $\nu$-dependent power-law exponent that varies from $2.3$ at $\nu = 0.4$ to $\sim 2.7$ at $\nu = -0.11$. 
While this is an interesting prospect, ideally one could resolve a larger scaling regime to reduce uncertainty in a potential power law. 
This might require higher resolution grains which can resolve distortions in shape of $\alpha < 10^{-3}$.

\begin{figure}
\begin{centering}	
\includegraphics[width=0.9\columnwidth]{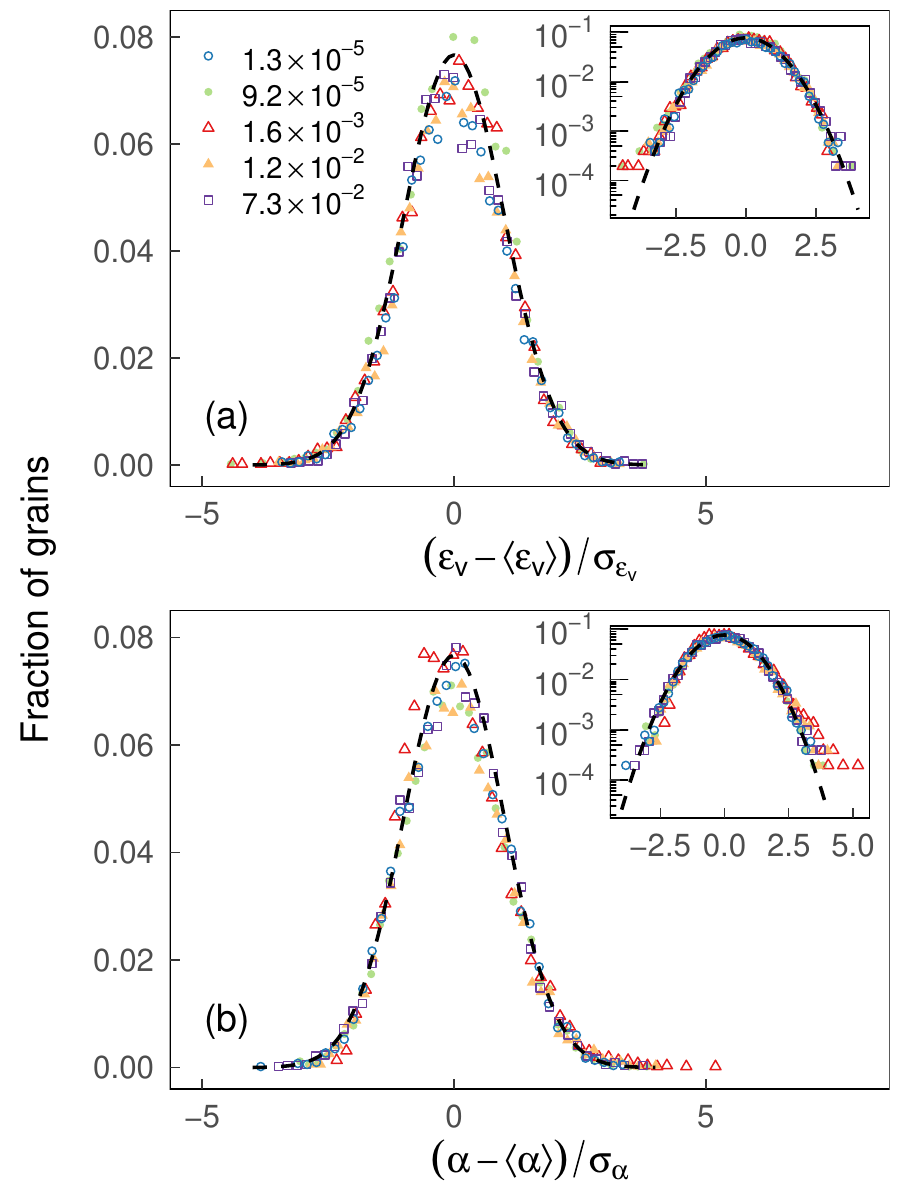}
	\caption{Distribution of the (a) volumetric strains and (b) shape factors of grains at $\nu = 1/4$ at the indicated values of $P/E_\mathrm{eff}$ after subtracting the mean and normalizing by the standard deviation (z-scores). A Gaussian curve is rendered as a black dashed line for comparison. Insets present the same data on a logarithmic scale.}
	\label{fig:grain_deformation_distribution}
\end{centering}
\end{figure}

Lastly, we consider the variation in granular deformation within a packing.
To predict yield and the accumulation of damage in powders, it is important to know the likelihood of a particular strain state in grains.
In all cases studied, rough vs. smooth particles or varying $\nu$, both the volumetric strain $\epsilon_v$ (Fig. \ref{fig:grain_deformation_distribution}[a]) and the asphericity $\alpha$ (Fig. \ref{fig:grain_deformation_distribution}[b]) of grains are approximately normally distributed.
As seen above, the averages grow with increasing pressure and the standard deviations exhibit similar growth with a consistent power law.
However, aside from the growth in these two values, the distributions exhibit a minimal dependence on pressure despite the significant differences in mechanical and scaling behavior seen in the low and high pressure limits.

%==============================================================%
\subsection{Elastic properties of the granular packing}
\label{sec:packedproperties}
%==============================================================%

\begin{figure}
\begin{centering}	
\includegraphics[width=0.9\columnwidth]{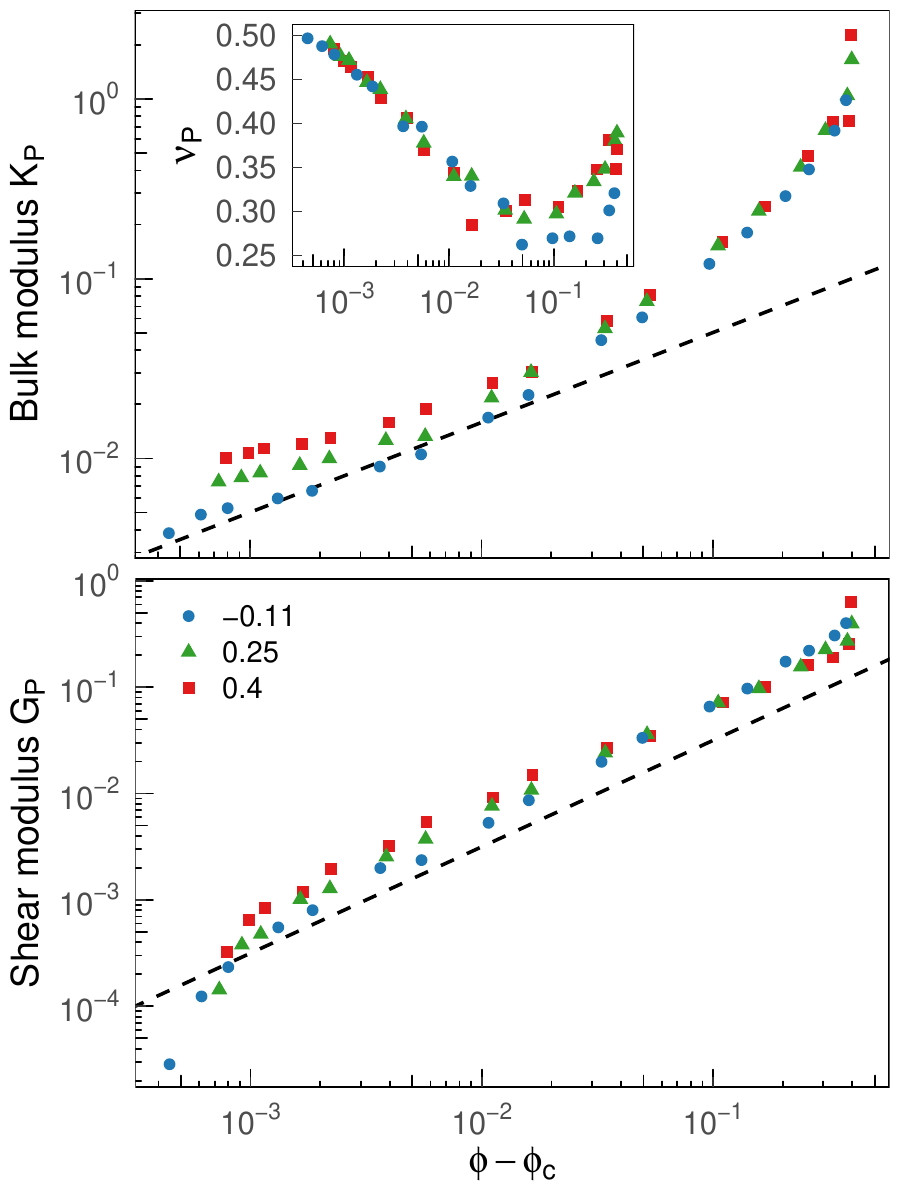}
	\caption{The (a) bulk and (b) shear moduli of the granular packing as a function of the distance to jamming for grains with the indicated values of $\nu$, the Poisson's ratio for individual grains. Dashed lines have slopes of (a) 1/2 and (b) 1. The inset plots the corresponding Poisson's ratios of the packing.}
	\label{fig:packing_moduli}
\end{centering}
\end{figure}

As a final study, we consider the elastic properties of the granular packing itself as the system densifies. 
Given the state of the BPM simulation at each value of $\phi$, brief loading experiments were performed by applying a small volumetric compression of 0.06\% strain or simple shear of 0.04\% strain at a slow rate to measure the stress response while simulating the particle dynamics. 
Linear regression is used to fit moduli from the stress response after excluding the first third of data to avoid potential minor shifts in particle positions resulting from an imperfect computational restarting procedure. 
Near the jamming transition, the bulk modulus of a disordered packing of Hertzian spheres $K_P$ is expected to grow as $(\phi - \phi_c)^{1/2}$. \cite{OHern2003, Agnolin2007}
For the shear modulus, there are subtleties and different power laws can be identified depending on whether particles can nonaffinely shift during the loading procedure.
As we allow particles to rearrange, we expect the shear modulus $G_P$ to grow proportional to $(\phi - \phi_c)$.\cite{OHern2003, Agnolin2007,Wang2021b}
This scaling is evident in Fig. \ref{fig:packing_moduli} where both $K_P$ and $G_P$ grow approximately according to their corresponding power law before accelerating at higher packing fractions.

Very close to jamming, $G_P$ is much smaller than $K_P$ such that the Poisson's ratio of the packing $\nu_P$ is 1/2, as seen in the inset of Fig. \ref{fig:moduli}(a).
This is true regardless of the Poisson's ratio of the constituent material $\nu$.
As the system densifies, the faster rise in $G_P$ causes $\nu_P$ to decrease. 
This continues until $\phi - \phi_c \sim 0.1$ where $\nu_P$ reaches a minimum value before rising.
Due to substantial variation in data, we cannot exactly determine the minimum of $\nu_P$ but it may interestingly reach $\sim 1/4$, particularly in the dataset with $\nu = -0.11$.
While this may simply be a coincidence, a nadir at $\nu_P = 1/4$ could reflect the limitations due to Cauchy's relations discussed in Sec. \ref{sec:bpm}.
At this intermediate pressure, relatively strong forces are expected to be exerted between particles such that one might no longer expect the contact topology to significantly change under small loads.
Furthermore due to the convergence of results between systems with different friction coefficients seen in Sec. \ref{sec:standard}, tangential frictional forces become less important such that forces could be effectively central-body.
Lastly, at even higher $\phi$ results began to diverge between BPM and DEM simulations implying contacts became non-Hertzian and may elastically interact with other contacts. 
This threshold may therefore mark the end of the assumption that grain-grain interactions are two-body.
Thus, this minimum could be related to a unique transitional density where $\nu_P \sim 1/4$ due to approximately pairwise central-body intergranular forces.
However, detailed characterization of grain-grain forces and more accurate measurements of elastic moduli are necessary.

As $\phi \rightarrow 1$, $\nu_P$ increases again due to a faster rise in the bulk modulus than the shear modulus. 
While we do know not exactly what causes this effect, it may be due to slip becoming relatively easier at grain-grain contacts as asperities flatten due to high normal forces decreasing $G_P/K_P$.
This type of an effect has also been previously observed in various experimental studies of compacted powders \cite{Carnavas1998, Hentschel2007}.
In particular, measurements by \citet{Hentschel2007} similarly found a minimum Poisson's ratio of around 1/4, or slightly higher, in various packed powders including copper, steel, aluminum, and glass.
Such characterizations are important in the development of state equations for compacted granular material and for predicting the strength of pressed particulate materials.

%==============================================================%
\section{Conclusion}
\label{sec:conclusion}
%==============================================================%

As many processes in industry and nature are far from the jamming threshold, there is a critical need to understand the high pressure compaction of granular materials.
While high pressures can be associated with elastic deformation, plasticity, and/or fracture, here we limited the problem to linear elasticity to quantify the effect of friction and elastic parameters.
To accomplish this task, we used a bonded particle model (BPM) where 1,024 spherical grains were each represented using $\sim 5,000$ particles.
The friction of grains was controlled via manipulating the morphology of surface particles, an approach which may mirror real-world processes of smoothing surfaces to control friction.
To vary the internal elasticity of grains, we proposed a new multibody interaction term which is relatively computationally cheap, easy to interpret, and can both decrease and increase Poisson's ratio $\nu$. 
This approach and the efficient implementation in LAMMPS allowed us to model a substantially larger sample of 3D grains than previous works on deformable grains in compaction \cite{Boromand2018, Boromand2019, Wang2021, Cardenas-Barrantes2022}.
 
At low pressures, results were consistent with the expectation that higher friction systems jam at lower critical packing fractions $\phi_c$ \cite{Silbert2010}.
Results from BPM simulations also obeyed the expected power-law scalings and agreed with results from traditional DEM simulations.
By a packing fraction of $\phi \sim \phi_c + 0.1$, we saw a crossover in several behaviors.
Compaction curves for systems with different friction coefficients began to converge onto a single curve, indicating frictional forces became less relevant.
The coalesced data also began to notably deviate from the power-law scaling seen near jamming implying grains have significantly deformed.
Similarly, a divergence between BPM and DEM results emerged at this threshold, indicating contact forces are no longer accurately described as Hertzian due to elastic interactions between contacts.
However, if one used analogous metrics, differences never exceeded a factor of two, suggesting one could still achieve qualitatively reasonable results using a considerably more economical DEM approach (assuming there is no nonlinear elasticity, plastic deformation, or fracture in the system).
Non-local DEM contact models that account for elastic interactions between contacts would likely improve DEM results \cite{Gonzalez2012, Giannis2021}.
The observed presence of a key crossover at $\phi \sim \phi_c + 0.1$ is also consistent with findings by \citet{Harthong2012} that the independence between contacts broke down near $\phi = 0.7$, close to $\phi-\phi_c = 0.1$, in MDFEM simulations of elastic-plastic grains.

Using the multibody term, we were able to vary the Poisson's ratio of grains $\nu$ from $0.4$ to $-0.11$.
While variations in $\nu$ did not affect standard metrics of pressure or coordination number with increasing packing fraction (beyond an analytic scaling due to the stiffness of contacts), significant differences were found in the deformation of grains.
Generally, the average volumetric compressive strain and the average asphericity of a grain, as studied in Refs. \cite{Boromand2018, Cardenas-Barrantes2022}, grow as a power of $\phi -\phi_c$ with approximate exponents of $3/2$ and $5/2$, respectively.
With increasing $\nu$, however, grains had less volumetric compression as curves shift downwards.
This result is expected but has not previously been quantified.

A more complex relationship between $\nu$ and asphericity was identified where the auxetic limit or $\nu < 0$ is associated with generally less distortion but a faster rate of growth in asphericity with increasing packing fraction.
This behavior could be described by a $\nu$-dependent power-law exponent that decreases from $2.7$ at $\nu = -0.11$ to $2.3$ at $\nu = 0.4$ and extends up to $\phi - \phi_c \sim 0.1$ before curves at different $\nu$ begin to converge as spheres inevitably have to distort to fully fill a volume.
Within the packing, the distribution of the volumetric strain and asphericity of grains remains approximately Gaussian at all pressures.
Such characterizations are important in predicting the probability of grains accumulating damage or fracturing as they strain with increasing load.

As the system densifies, interesting trends in the elastic properties of the packing also emerge.
Near jamming, the shear modulus of the packed system grows faster than the bulk modulus with increasing packing fraction as previously noted in the literature \cite{OHern2003, Agnolin2007,Wang2021b}.
This leads to a reduction in the Poisson's ratio of the packed system $\nu_P$ that continues until $\phi- \phi_c \sim 0.1$, another indication of an important transition in the compaction of grains.
At larger $\phi$, the bulk modulus then grows faster than the shear modulus causing $\nu_P$ to rise.
This creates a minimal value of $\nu_P$ that may equal $1/4$ and could reflect the unique nature of grain-grain interactions at this transition between the hard-particle and a deformation-dominated limit.
The presence of a minimum value of $\nu_P$ has been seen before in experimental studies \cite{Carnavas1998, Hentschel2007}.

Despite the breadth of topics explored, this work still only touches upon an idealized limit.
For instance, real materials are not monodisperse, defect-free spheres and non-linearities in elasticity are expected to emerge at large deformations.
Furthermore, many applications depend on the activation of inelastic mechanisms not modeled here.
However, such characteristics could be captured using the BPM tools presented here offering a wide scope for future work.

%==============================================================%

\section*{Author Contributions}
\textbf{Joel T. Clemmer}: Formal analysis; conceptualization; software; writing - original draft.
\textbf{Joseph M. Monti}: Formal analysis; writing - review \& editing.
\textbf{Jeremy B. Lechman}: Conceptualization; funding acquisition; writing - review \& editing.

\section*{Conflicts of interest}
There are no conflicts to declare.

\section*{Acknowledgements}
We thank D. S. Bolintineanu, G. S. Grest, and L. E. Silbert for useful discussions.
This work is funded by the Advanced Simulation and Computing program.
This article has been authored by an employee of National Technology \& Engineering Solutions of Sandia, LLC under Contract No. DE-NA0003525 with the U.S. Department of Energy (DOE). The employee owns all right, title and interest in and to the article and is solely responsible for its contents. The United States Government retains and the publisher, by accepting the article for publication, acknowledges that the United States Government retains a non-exclusive, paid-up, irrevocable, world-wide license to publish or reproduce the published form of this article or allow others to do so, for United States Government purposes. The DOE will provide public access to these results of federally sponsored research in accordance with the DOE Public Access Plan https://www.energy.gov/downloads/doe-public-access-plan.
This paper describes objective technical results and analysis. Any subjective views or opinions that might be expressed in the paper do not necessarily represent the views of the U.S. Department of Energy or the United States Government.

\appendix

%==============================================================%
\section{Convergence properties of the BPM}
\label{sec:convergence}
%==============================================================%

An effective computational model of continuum elasticity must converge to the true analytic solution with increasing simulation resolution.
To quantify convergence properties of our BPM simulations, we tested the response of different sized systems to homogeneous deformation.
Here we use the same isotropic compression and simple shear deformation geometries used to calibrate material properties in Sec. \ref{sec:calibrate} but vary the linear size of the system from $L = 2$ to $200$ particle diameters or $\sim 10$ to $10^7$ particles.
Multiple random realizations are created for each system size (except for $L = 200$ which uses a single realization) and the set of measured bulk moduli $K$ are then used to calculate an average value $K(L)$ and a standard deviation $\sigma_K(L)$ at each system size.

\begin{figure}
\begin{centering}
	\includegraphics[width=0.9\columnwidth]{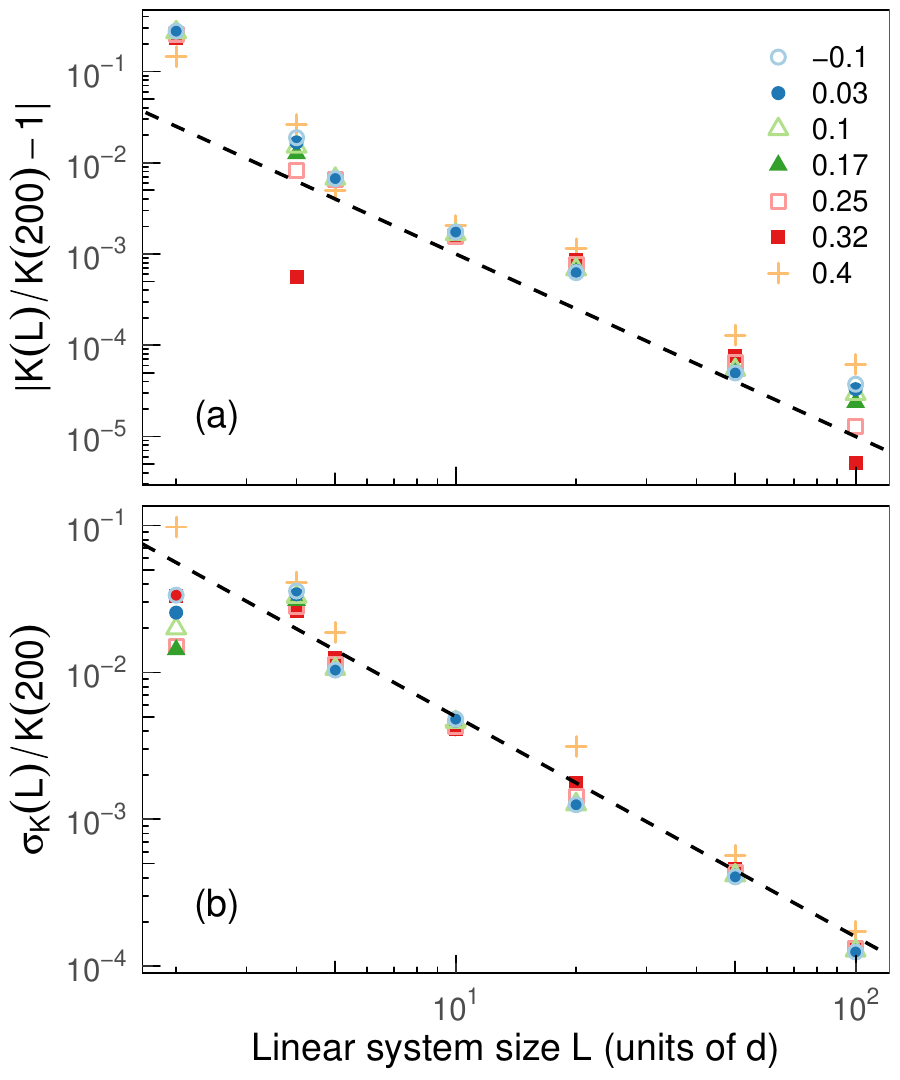}
	\caption{(a) The average deviation in the bulk modulus $K$ at a system size $L$ relative to a system of size $L = 200$, or $K(200)$, for the indicated Poisson's ratios. (b) The standard deviation of measured $K$ normalized by $K(200)$ as a function of $L$. Dashed lines have slopes of (a) $-2$ and (b) $-3/2$.}
	\label{fig:conv_moduli}
\end{centering}
\end{figure}

Assuming $K(200)$ is representative of an infinite sized system, the deviation in $K(L)$ from $K(200)$ as a function of $L$ is plotted in Fig. \ref{fig:conv_moduli}(a).
As $L$ increases, the deviation decays as $L^{-2}$ across all values of $a_B$ or $\nu$.
In Fig. \ref{fig:conv_moduli}(b), $\sigma_K(L)$ decays as $L^{-3/2}$.
This power-law could emerge from an interpretation of the bulk modulus of a large system as the cumulative sum of $N\sim L^3$ independent random bulk moduli measured in smaller regions.
Very similar results are seen for the shear modulus $G$.
This implies the elastic properties of the BPM systems converge to a limiting value.
For systems of size $L \sim 16$ with $\sim 5,000$ particles, similar to the size of single grain, we expect variations of less than 1\% in material properties due to the random packing of particles. 

\begin{figure}
\begin{centering}
	\includegraphics[width=0.9\columnwidth]{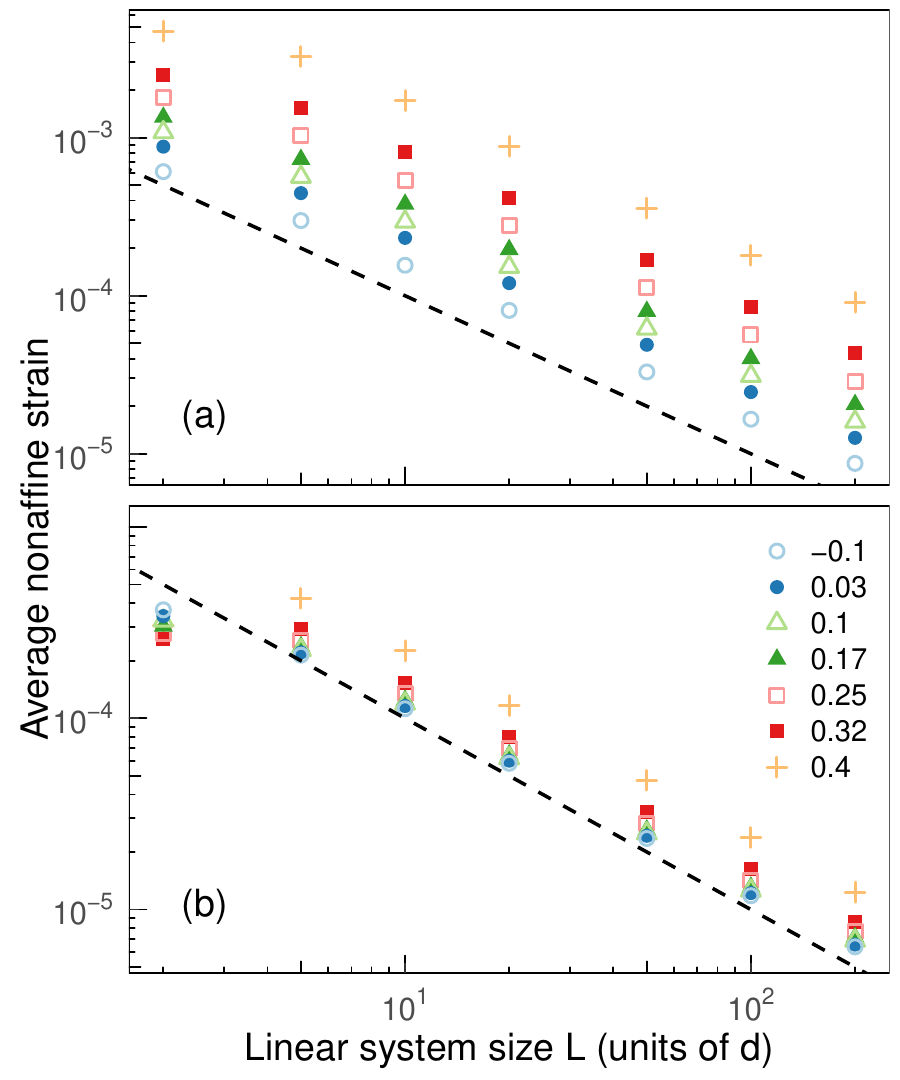}
	\caption{The average nonaffine strain of particles against system size $L$ in (a) compression and (b) simple shear for the indicated Poisson's ratios. Dashed lines have slopes of (a) $-1$.}
	\label{fig:conv_disp}
\end{centering}
\end{figure}

As an additional check, we also calculated the average nonaffine strains of particles after $2.5\%$ volumetric strain under compression or $2.5\%$ shear strain under simple shear plotted in Fig. \ref{fig:conv_disp}(a) and (b), respectively.
This quantifies how much the local strain field in a system deviates from the true elastic solution.
A nonaffine displacement of each particle is essentially how far it moves due to unbalanced forces that emerge as the system deforms.
A strain is then defined by normalizing the displacement by the system size.
Since the nonaffine displacement of a particle is effectively independent of the system size and is just a property of the initial packing, the average nonaffine strains simply decays as $1/L$. 
One may be able to adjust the magnitude of this effect by controlling the quench rate of the initial packing which is associated with the local yield strength of glasses \cite{Patinet2016}.

While the nonaffine strain decays with the same exponent in both compression and simple shear across all Poisson's ratios  $\nu$ simulated, there is a vertical shift associated with the specific value of $\nu$.
Particles have larger nonaffine strains near the incompressible limit at large $\nu$.
This effect is quite prominent in compression but is relatively minor in simple shear.
This likely stems from the ability for particles to rearrange relative to their neighbors to optimize the distribution of local volumes to minimize forces generated by the multibody term at small $a_B$, associated with the dependence of $K$ on $a_B$ noted in Fig. \ref{fig:moduli}(b).
As postulated in the methods, the magnitude of this effect might be reduced if the multibody term was constrained to apply a net force of zero across all of a particle's bonds.

\bibliography{rsc}

\providecommand*{\mcitethebibliography}{\thebibliography}
\csname @ifundefined\endcsname{endmcitethebibliography}
{\let\endmcitethebibliography\endthebibliography}{}
\begin{mcitethebibliography}{76}
\providecommand*{\natexlab}[1]{#1}
\providecommand*{\mciteSetBstSublistMode}[1]{}
\providecommand*{\mciteSetBstMaxWidthForm}[2]{}
\providecommand*{\mciteBstWouldAddEndPuncttrue}
  {\def\EndOfBibitem{\unskip.}}
\providecommand*{\mciteBstWouldAddEndPunctfalse}
  {\let\EndOfBibitem\relax}
\providecommand*{\mciteSetBstMidEndSepPunct}[3]{}
\providecommand*{\mciteSetBstSublistLabelBeginEnd}[3]{}
\providecommand*{\EndOfBibitem}{}
\mciteSetBstSublistMode{f}
\mciteSetBstMaxWidthForm{subitem}
{(\emph{\alph{mcitesubitemcount}})}
\mciteSetBstSublistLabelBeginEnd{\mcitemaxwidthsubitemform\space}
{\relax}{\relax}

\bibitem[Torquato and Stillinger(2010)]{Torquato2010}
S.~Torquato and F.~H. Stillinger, \emph{Rev. Mod. Phys.}, 2010, \textbf{82},
  2633--2672\relax
\mciteBstWouldAddEndPuncttrue
\mciteSetBstMidEndSepPunct{\mcitedefaultmidpunct}
{\mcitedefaultendpunct}{\mcitedefaultseppunct}\relax
\EndOfBibitem
\bibitem[van Hecke(2010)]{VanHecke2010}
M.~van Hecke, \emph{J. Phys. Condens. Matter}, 2010, \textbf{22}, 033101\relax
\mciteBstWouldAddEndPuncttrue
\mciteSetBstMidEndSepPunct{\mcitedefaultmidpunct}
{\mcitedefaultendpunct}{\mcitedefaultseppunct}\relax
\EndOfBibitem
\bibitem[Behringer and Chakraborty(2019)]{Behringer2019}
R.~P. Behringer and B.~Chakraborty, \emph{Reports Prog. Phys.}, 2019,
  \textbf{82}, 012601\relax
\mciteBstWouldAddEndPuncttrue
\mciteSetBstMidEndSepPunct{\mcitedefaultmidpunct}
{\mcitedefaultendpunct}{\mcitedefaultseppunct}\relax
\EndOfBibitem
\bibitem[Silbert(2010)]{Silbert2010}
L.~E. Silbert, \emph{Soft Matter}, 2010, \textbf{6}, 2918\relax
\mciteBstWouldAddEndPuncttrue
\mciteSetBstMidEndSepPunct{\mcitedefaultmidpunct}
{\mcitedefaultendpunct}{\mcitedefaultseppunct}\relax
\EndOfBibitem
\bibitem[Santos \emph{et~al.}(2020)Santos, Bolintineanu, Grest, Lechman,
  Plimpton, Srivastava, and Silbert]{Santos2020}
A.~P. Santos, D.~S. Bolintineanu, G.~S. Grest, J.~B. Lechman, S.~J. Plimpton,
  I.~Srivastava and L.~E. Silbert, \emph{Phys. Rev. E}, 2020, \textbf{102},
  032903\relax
\mciteBstWouldAddEndPuncttrue
\mciteSetBstMidEndSepPunct{\mcitedefaultmidpunct}
{\mcitedefaultendpunct}{\mcitedefaultseppunct}\relax
\EndOfBibitem
\bibitem[Erikson \emph{et~al.}(2002)Erikson, Mueggenburg, Jaeger, and
  Nagel]{Erikson2002}
J.~M. Erikson, N.~W. Mueggenburg, H.~M. Jaeger and S.~R. Nagel, \emph{Phys.
  Rev. E - Stat. Physics, Plasmas, Fluids, Relat. Interdiscip. Top.}, 2002,
  \textbf{66}, 4\relax
\mciteBstWouldAddEndPuncttrue
\mciteSetBstMidEndSepPunct{\mcitedefaultmidpunct}
{\mcitedefaultendpunct}{\mcitedefaultseppunct}\relax
\EndOfBibitem
\bibitem[Brodu \emph{et~al.}(2015)Brodu, Dijksman, and Behringer]{Brodu2015}
N.~Brodu, J.~A. Dijksman and R.~P. Behringer, \emph{Nat. Commun.}, 2015,
  \textbf{6}, 6361\relax
\mciteBstWouldAddEndPuncttrue
\mciteSetBstMidEndSepPunct{\mcitedefaultmidpunct}
{\mcitedefaultendpunct}{\mcitedefaultseppunct}\relax
\EndOfBibitem
\bibitem[Vu \emph{et~al.}(2020)Vu, Nezamabadi, and Mora]{Vu2020}
T.-L. Vu, S.~Nezamabadi and S.~Mora, \emph{Soft Matter}, 2020, \textbf{16},
  679--687\relax
\mciteBstWouldAddEndPuncttrue
\mciteSetBstMidEndSepPunct{\mcitedefaultmidpunct}
{\mcitedefaultendpunct}{\mcitedefaultseppunct}\relax
\EndOfBibitem
\bibitem[Bar{\'{e}}s \emph{et~al.}(2022)Bar{\'{e}}s, C{\'{a}}rdenas-Barrantes,
  Cantor, Renouf, and Az{\'{e}}ma]{Bares2022}
J.~Bar{\'{e}}s, M.~C{\'{a}}rdenas-Barrantes, D.~Cantor, M.~Renouf and
  {\'{E}}.~Az{\'{e}}ma, \emph{Pap. Phys.}, 2022, \textbf{14}, 140009\relax
\mciteBstWouldAddEndPuncttrue
\mciteSetBstMidEndSepPunct{\mcitedefaultmidpunct}
{\mcitedefaultendpunct}{\mcitedefaultseppunct}\relax
\EndOfBibitem
\bibitem[Bar{\'{e}}s \emph{et~al.}(2023)Bar{\'{e}}s, C{\'{a}}rdenas-Barrantes,
  Pinz{\'{o}}n, And{\'{o}}, Renouf, Viggiani, and Az{\'{e}}ma]{Bares2023}
J.~Bar{\'{e}}s, M.~C{\'{a}}rdenas-Barrantes, G.~Pinz{\'{o}}n, E.~And{\'{o}},
  M.~Renouf, G.~Viggiani and E.~Az{\'{e}}ma, 2023,  1--24\relax
\mciteBstWouldAddEndPuncttrue
\mciteSetBstMidEndSepPunct{\mcitedefaultmidpunct}
{\mcitedefaultendpunct}{\mcitedefaultseppunct}\relax
\EndOfBibitem
\bibitem[Gethin \emph{et~al.}(2003)Gethin, Lewis, and Ransing]{Gethin2003}
D.~T. Gethin, R.~W. Lewis and R.~S. Ransing, \emph{Model. Simul. Mater. Sci.
  Eng.}, 2003, \textbf{11}, 101--114\relax
\mciteBstWouldAddEndPuncttrue
\mciteSetBstMidEndSepPunct{\mcitedefaultmidpunct}
{\mcitedefaultendpunct}{\mcitedefaultseppunct}\relax
\EndOfBibitem
\bibitem[Procopio and Zavaliangos(2005)]{Procopio2005}
A.~T. Procopio and A.~Zavaliangos, \emph{J. Mech. Phys. Solids}, 2005,
  \textbf{53}, 1523--1551\relax
\mciteBstWouldAddEndPuncttrue
\mciteSetBstMidEndSepPunct{\mcitedefaultmidpunct}
{\mcitedefaultendpunct}{\mcitedefaultseppunct}\relax
\EndOfBibitem
\bibitem[Cantor \emph{et~al.}(2020)Cantor, C{\'{a}}rdenas-Barrantes,
  Preechawuttipong, Renouf, and Az{\'{e}}ma]{Cantor2020}
D.~Cantor, M.~C{\'{a}}rdenas-Barrantes, I.~Preechawuttipong, M.~Renouf and
  E.~Az{\'{e}}ma, \emph{Phys. Rev. Lett.}, 2020, \textbf{124}, 2--7\relax
\mciteBstWouldAddEndPuncttrue
\mciteSetBstMidEndSepPunct{\mcitedefaultmidpunct}
{\mcitedefaultendpunct}{\mcitedefaultseppunct}\relax
\EndOfBibitem
\bibitem[C{\'{a}}rdenas-Barrantes \emph{et~al.}(2022)C{\'{a}}rdenas-Barrantes,
  Cantor, Bar{\'{e}}s, Renouf, and Az{\'{e}}ma]{Cardenas-Barrantes2022}
M.~C{\'{a}}rdenas-Barrantes, D.~Cantor, J.~Bar{\'{e}}s, M.~Renouf and
  E.~Az{\'{e}}ma, \emph{Soft Matter}, 2022, \textbf{18}, 312--321\relax
\mciteBstWouldAddEndPuncttrue
\mciteSetBstMidEndSepPunct{\mcitedefaultmidpunct}
{\mcitedefaultendpunct}{\mcitedefaultseppunct}\relax
\EndOfBibitem
\bibitem[Nezamabadi \emph{et~al.}(2017)Nezamabadi, Nguyen, Delenne, and
  Radjai]{Nezamabadi2017}
S.~Nezamabadi, T.~H. Nguyen, J.-Y. Delenne and F.~Radjai, \emph{Granul.
  Matter}, 2017, \textbf{19}, 8\relax
\mciteBstWouldAddEndPuncttrue
\mciteSetBstMidEndSepPunct{\mcitedefaultmidpunct}
{\mcitedefaultendpunct}{\mcitedefaultseppunct}\relax
\EndOfBibitem
\bibitem[Vu \emph{et~al.}(2021)Vu, Nezamabadi, and Mora]{Vu2021}
T.~L. Vu, S.~Nezamabadi and S.~Mora, \emph{J. Mech. Phys. Solids}, 2021,
  \textbf{146}, 104201\relax
\mciteBstWouldAddEndPuncttrue
\mciteSetBstMidEndSepPunct{\mcitedefaultmidpunct}
{\mcitedefaultendpunct}{\mcitedefaultseppunct}\relax
\EndOfBibitem
\bibitem[Dosta \emph{et~al.}(2017)Dosta, Costa, and Al-Qureshi]{Dosta2017}
M.~Dosta, C.~Costa and H.~Al-Qureshi, \emph{EPJ Web Conf.}, 2017, \textbf{140},
  15021\relax
\mciteBstWouldAddEndPuncttrue
\mciteSetBstMidEndSepPunct{\mcitedefaultmidpunct}
{\mcitedefaultendpunct}{\mcitedefaultseppunct}\relax
\EndOfBibitem
\bibitem[Giannis \emph{et~al.}(2023)Giannis, Kwade, Finke, and
  Schilde]{Giannis2023}
K.~Giannis, A.~Kwade, J.~H. Finke and C.~Schilde, \emph{Pharmaceutics}, 2023,
  \textbf{15}, 909\relax
\mciteBstWouldAddEndPuncttrue
\mciteSetBstMidEndSepPunct{\mcitedefaultmidpunct}
{\mcitedefaultendpunct}{\mcitedefaultseppunct}\relax
\EndOfBibitem
\bibitem[Boromand \emph{et~al.}(2018)Boromand, Signoriello, Ye, O'Hern, and
  Shattuck]{Boromand2018}
A.~Boromand, A.~Signoriello, F.~Ye, C.~S. O'Hern and M.~D. Shattuck,
  \emph{Phys. Rev. Lett.}, 2018, \textbf{121}, 248003\relax
\mciteBstWouldAddEndPuncttrue
\mciteSetBstMidEndSepPunct{\mcitedefaultmidpunct}
{\mcitedefaultendpunct}{\mcitedefaultseppunct}\relax
\EndOfBibitem
\bibitem[Boromand \emph{et~al.}(2019)Boromand, Signoriello, Lowensohn,
  Orellana, Weeks, Ye, Shattuck, and O'Hern]{Boromand2019}
A.~Boromand, A.~Signoriello, J.~Lowensohn, C.~S. Orellana, E.~R. Weeks, F.~Ye,
  M.~D. Shattuck and C.~S. O'Hern, \emph{Soft Matter}, 2019, \textbf{15},
  5854--5865\relax
\mciteBstWouldAddEndPuncttrue
\mciteSetBstMidEndSepPunct{\mcitedefaultmidpunct}
{\mcitedefaultendpunct}{\mcitedefaultseppunct}\relax
\EndOfBibitem
\bibitem[Wang \emph{et~al.}(2021)Wang, Treado, Boromand, Norwick, Murrell,
  Shattuck, and O'Hern]{Wang2021}
D.~Wang, J.~D. Treado, A.~Boromand, B.~Norwick, M.~P. Murrell, M.~D. Shattuck
  and C.~S. O'Hern, \emph{Soft Matter}, 2021, \textbf{17}, 9901--9915\relax
\mciteBstWouldAddEndPuncttrue
\mciteSetBstMidEndSepPunct{\mcitedefaultmidpunct}
{\mcitedefaultendpunct}{\mcitedefaultseppunct}\relax
\EndOfBibitem
\bibitem[Goodrich \emph{et~al.}(2015)Goodrich, Liu, and Nagel]{Goodrich2015}
C.~P. Goodrich, A.~J. Liu and S.~R. Nagel, \emph{Phys. Rev. Lett.}, 2015,
  \textbf{114}, 225501\relax
\mciteBstWouldAddEndPuncttrue
\mciteSetBstMidEndSepPunct{\mcitedefaultmidpunct}
{\mcitedefaultendpunct}{\mcitedefaultseppunct}\relax
\EndOfBibitem
\bibitem[Santos \emph{et~al.}(2022)Santos, Srivastava, Silbert, Lechman, and
  Grest]{Santos2022}
A.~P. Santos, I.~Srivastava, L.~E. Silbert, J.~B. Lechman and G.~S. Grest,
  \emph{Phys. Rev. Fluids}, 2022, \textbf{7}, 084303\relax
\mciteBstWouldAddEndPuncttrue
\mciteSetBstMidEndSepPunct{\mcitedefaultmidpunct}
{\mcitedefaultendpunct}{\mcitedefaultseppunct}\relax
\EndOfBibitem
\bibitem[Thompson \emph{et~al.}(2022)Thompson, Aktulga, Berger, Bolintineanu,
  Brown, Crozier, {in 't Veld}, Kohlmeyer, Moore, Nguyen, Shan, Stevens,
  Tranchida, Trott, and Plimpton]{Thompson2021}
A.~P. Thompson, H.~M. Aktulga, R.~Berger, D.~S. Bolintineanu, W.~M. Brown,
  P.~S. Crozier, P.~J. {in 't Veld}, A.~Kohlmeyer, S.~G. Moore, T.~D. Nguyen,
  R.~Shan, M.~J. Stevens, J.~Tranchida, C.~Trott and S.~J. Plimpton,
  \emph{Comput. Phys. Commun.}, 2022, \textbf{271}, 108171\relax
\mciteBstWouldAddEndPuncttrue
\mciteSetBstMidEndSepPunct{\mcitedefaultmidpunct}
{\mcitedefaultendpunct}{\mcitedefaultseppunct}\relax
\EndOfBibitem
\bibitem[Cundall and Strack(1979)]{Cundall1979}
P.~A. Cundall and O.~D.~L. Strack, \emph{Géotechnique}, 1979, \textbf{29},
  47--65\relax
\mciteBstWouldAddEndPuncttrue
\mciteSetBstMidEndSepPunct{\mcitedefaultmidpunct}
{\mcitedefaultendpunct}{\mcitedefaultseppunct}\relax
\EndOfBibitem
\bibitem[Gonzalez and Cuiti{\~{n}}o(2012)]{Gonzalez2012}
M.~Gonzalez and A.~M. Cuiti{\~{n}}o, \emph{J. Mech. Phys. Solids}, 2012,
  \textbf{60}, 333--350\relax
\mciteBstWouldAddEndPuncttrue
\mciteSetBstMidEndSepPunct{\mcitedefaultmidpunct}
{\mcitedefaultendpunct}{\mcitedefaultseppunct}\relax
\EndOfBibitem
\bibitem[Harthong \emph{et~al.}(2012)Harthong, J{\'{e}}rier, Richefeu,
  Chareyre, Dor{\'{e}}mus, Imbault, and Donz{\'{e}}]{Harthong2012}
B.~Harthong, J.-F. J{\'{e}}rier, V.~Richefeu, B.~Chareyre, P.~Dor{\'{e}}mus,
  D.~Imbault and F.-V. Donz{\'{e}}, \emph{Int. J. Mech. Sci.}, 2012,
  \textbf{61}, 32--43\relax
\mciteBstWouldAddEndPuncttrue
\mciteSetBstMidEndSepPunct{\mcitedefaultmidpunct}
{\mcitedefaultendpunct}{\mcitedefaultseppunct}\relax
\EndOfBibitem
\bibitem[Behzadinasab \emph{et~al.}(2018)Behzadinasab, Vogler, Peterson,
  Rahman, and Foster]{Behzadinasab2018}
M.~Behzadinasab, T.~J. Vogler, A.~M. Peterson, R.~Rahman and J.~T. Foster,
  \emph{J. Dyn. Behav. Mater.}, 2018, \textbf{4}, 529--542\relax
\mciteBstWouldAddEndPuncttrue
\mciteSetBstMidEndSepPunct{\mcitedefaultmidpunct}
{\mcitedefaultendpunct}{\mcitedefaultseppunct}\relax
\EndOfBibitem
\bibitem[Silling \emph{et~al.}(2021)Silling, Barr, Cooper, Lechman, and
  Bufford]{Silling2021}
S.~A. Silling, C.~Barr, M.~Cooper, J.~Lechman and D.~C. Bufford, \emph{Comput.
  Part. Mech.}, 2021, \textbf{8}, 1005--1017\relax
\mciteBstWouldAddEndPuncttrue
\mciteSetBstMidEndSepPunct{\mcitedefaultmidpunct}
{\mcitedefaultendpunct}{\mcitedefaultseppunct}\relax
\EndOfBibitem
\bibitem[Homel and Herbold(2017)]{Homel2017}
M.~A. Homel and E.~B. Herbold, \emph{Int. J. Numer. Methods Eng.}, 2017,
  \textbf{109}, 1013--1044\relax
\mciteBstWouldAddEndPuncttrue
\mciteSetBstMidEndSepPunct{\mcitedefaultmidpunct}
{\mcitedefaultendpunct}{\mcitedefaultseppunct}\relax
\EndOfBibitem
\bibitem[Lisjak and Grasselli(2014)]{Lisjak2014}
A.~Lisjak and G.~Grasselli, \emph{J. Rock Mech. Geotech. Eng.}, 2014,
  \textbf{6}, 301--314\relax
\mciteBstWouldAddEndPuncttrue
\mciteSetBstMidEndSepPunct{\mcitedefaultmidpunct}
{\mcitedefaultendpunct}{\mcitedefaultseppunct}\relax
\EndOfBibitem
\bibitem[Potyondy and Cundall(2004)]{Potyondy2004}
D.~O. Potyondy and P.~A. Cundall, \emph{Int. J. Rock Mech. Min. Sci.}, 2004,
  \textbf{41}, 1329--1364\relax
\mciteBstWouldAddEndPuncttrue
\mciteSetBstMidEndSepPunct{\mcitedefaultmidpunct}
{\mcitedefaultendpunct}{\mcitedefaultseppunct}\relax
\EndOfBibitem
\bibitem[Andr{\'{e}} \emph{et~al.}(2012)Andr{\'{e}}, Iordanoff, Charles, and
  N{\'{e}}auport]{Andre2012}
D.~Andr{\'{e}}, I.~Iordanoff, J.-l. Charles and J.~N{\'{e}}auport,
  \emph{Comput. Methods Appl. Mech. Eng.}, 2012, \textbf{213-216},
  113--125\relax
\mciteBstWouldAddEndPuncttrue
\mciteSetBstMidEndSepPunct{\mcitedefaultmidpunct}
{\mcitedefaultendpunct}{\mcitedefaultseppunct}\relax
\EndOfBibitem
\bibitem[Celigueta \emph{et~al.}(2017)Celigueta, Latorre, Arrufat, and
  O{\~{n}}ate]{Celigueta2017}
M.~A. Celigueta, S.~Latorre, F.~Arrufat and E.~O{\~{n}}ate, \emph{Comput.
  Mech.}, 2017, \textbf{60}, 997--1010\relax
\mciteBstWouldAddEndPuncttrue
\mciteSetBstMidEndSepPunct{\mcitedefaultmidpunct}
{\mcitedefaultendpunct}{\mcitedefaultseppunct}\relax
\EndOfBibitem
\bibitem[Ostoja-Starzewski(2002)]{Ostoja-Starzewski2002}
M.~Ostoja-Starzewski, \emph{Appl. Mech. Rev.}, 2002, \textbf{55}, 35--59\relax
\mciteBstWouldAddEndPuncttrue
\mciteSetBstMidEndSepPunct{\mcitedefaultmidpunct}
{\mcitedefaultendpunct}{\mcitedefaultseppunct}\relax
\EndOfBibitem
\bibitem[Cusatis \emph{et~al.}(2011)Cusatis, Pelessone, and
  Mencarelli]{Cusatis2011}
G.~Cusatis, D.~Pelessone and A.~Mencarelli, \emph{Cem. Concr. Compos.}, 2011,
  \textbf{33}, 881--890\relax
\mciteBstWouldAddEndPuncttrue
\mciteSetBstMidEndSepPunct{\mcitedefaultmidpunct}
{\mcitedefaultendpunct}{\mcitedefaultseppunct}\relax
\EndOfBibitem
\bibitem[Zhao \emph{et~al.}(2011)Zhao, Fang, and Zhao]{Zhao2011}
G.-F. Zhao, J.~Fang and J.~Zhao, \emph{Int. J. Numer. Anal. Methods Geomech.},
  2011, \textbf{35}, 859--885\relax
\mciteBstWouldAddEndPuncttrue
\mciteSetBstMidEndSepPunct{\mcitedefaultmidpunct}
{\mcitedefaultendpunct}{\mcitedefaultseppunct}\relax
\EndOfBibitem
\bibitem[Chen \emph{et~al.}(2014)Chen, Lin, Jiao, and Liu]{Chen2014}
H.~Chen, E.~Lin, Y.~Jiao and Y.~Liu, \emph{Comput. Mech.}, 2014, \textbf{54},
  1541--1558\relax
\mciteBstWouldAddEndPuncttrue
\mciteSetBstMidEndSepPunct{\mcitedefaultmidpunct}
{\mcitedefaultendpunct}{\mcitedefaultseppunct}\relax
\EndOfBibitem
\bibitem[Chen and Liu(2016)]{Chen2016}
H.~Chen and Y.~Liu, \emph{Int. J. Solids Struct.}, 2016, \textbf{81},
  411--420\relax
\mciteBstWouldAddEndPuncttrue
\mciteSetBstMidEndSepPunct{\mcitedefaultmidpunct}
{\mcitedefaultendpunct}{\mcitedefaultseppunct}\relax
\EndOfBibitem
\bibitem[Kot and Nagahashi(2017)]{Kot2017}
M.~Kot and H.~Nagahashi, \emph{Vis. Comput.}, 2017, \textbf{33}, 283--291\relax
\mciteBstWouldAddEndPuncttrue
\mciteSetBstMidEndSepPunct{\mcitedefaultmidpunct}
{\mcitedefaultendpunct}{\mcitedefaultseppunct}\relax
\EndOfBibitem
\bibitem[Golec \emph{et~al.}(2020)Golec, Palierne, Zara, Nicolle, and
  Damiand]{Golec2020}
K.~Golec, J.-F. Palierne, F.~Zara, S.~Nicolle and G.~Damiand, \emph{Vis.
  Comput.}, 2020, \textbf{36}, 809--825\relax
\mciteBstWouldAddEndPuncttrue
\mciteSetBstMidEndSepPunct{\mcitedefaultmidpunct}
{\mcitedefaultendpunct}{\mcitedefaultseppunct}\relax
\EndOfBibitem
\bibitem[Silling and Lehoucq(2010)]{Silling2010}
S.~Silling and R.~Lehoucq, in \emph{Adv. Appl. Mech.}, Elsevier, 2010, vol.~44,
  pp. 73--168\relax
\mciteBstWouldAddEndPuncttrue
\mciteSetBstMidEndSepPunct{\mcitedefaultmidpunct}
{\mcitedefaultendpunct}{\mcitedefaultseppunct}\relax
\EndOfBibitem
\bibitem[Trageser and Seleson(2020)]{Trageser2020}
J.~Trageser and P.~Seleson, \emph{J. Peridynamics Nonlocal Model.}, 2020,
  \textbf{2}, 278--288\relax
\mciteBstWouldAddEndPuncttrue
\mciteSetBstMidEndSepPunct{\mcitedefaultmidpunct}
{\mcitedefaultendpunct}{\mcitedefaultseppunct}\relax
\EndOfBibitem
\bibitem[Wang and Mora(2009)]{Wang2009b}
Y.~Wang and P.~Mora, in \emph{Adv. Geocomputing}, Springer Berlin Heidelberg,
  2009, vol. 119 of Lecture Notes in Earth Sciences, pp. 183--228\relax
\mciteBstWouldAddEndPuncttrue
\mciteSetBstMidEndSepPunct{\mcitedefaultmidpunct}
{\mcitedefaultendpunct}{\mcitedefaultseppunct}\relax
\EndOfBibitem
\bibitem[Tim{\'{a}}r and Kun(2011)]{Timar2011}
G.~Tim{\'{a}}r and F.~Kun, \emph{Phys. Rev. E}, 2011, \textbf{83}, 046115\relax
\mciteBstWouldAddEndPuncttrue
\mciteSetBstMidEndSepPunct{\mcitedefaultmidpunct}
{\mcitedefaultendpunct}{\mcitedefaultseppunct}\relax
\EndOfBibitem
\bibitem[Andr{\'{e}} \emph{et~al.}(2019)Andr{\'{e}}, Girardot, and
  Hubert]{Andre2019}
D.~Andr{\'{e}}, J.~Girardot and C.~Hubert, \emph{Comput. Methods Appl. Mech.
  Eng.}, 2019, \textbf{350}, 100--122\relax
\mciteBstWouldAddEndPuncttrue
\mciteSetBstMidEndSepPunct{\mcitedefaultmidpunct}
{\mcitedefaultendpunct}{\mcitedefaultseppunct}\relax
\EndOfBibitem
\bibitem[Carmona \emph{et~al.}(2008)Carmona, Wittel, Kun, and
  Herrmann]{Carmona2008}
H.~A. Carmona, F.~K. Wittel, F.~Kun and H.~J. Herrmann, \emph{Phys. Rev. E},
  2008, \textbf{77}, 051302\relax
\mciteBstWouldAddEndPuncttrue
\mciteSetBstMidEndSepPunct{\mcitedefaultmidpunct}
{\mcitedefaultendpunct}{\mcitedefaultseppunct}\relax
\EndOfBibitem
\bibitem[Wang(2009)]{Wang2009}
Y.~Wang, \emph{Acta Geotech.}, 2009, \textbf{4}, 117--127\relax
\mciteBstWouldAddEndPuncttrue
\mciteSetBstMidEndSepPunct{\mcitedefaultmidpunct}
{\mcitedefaultendpunct}{\mcitedefaultseppunct}\relax
\EndOfBibitem
\bibitem[Beale and Srolovitz(1988)]{Beale1988}
P.~D. Beale and D.~J. Srolovitz, \emph{Phys. Rev. B}, 1988, \textbf{37},
  5500--5507\relax
\mciteBstWouldAddEndPuncttrue
\mciteSetBstMidEndSepPunct{\mcitedefaultmidpunct}
{\mcitedefaultendpunct}{\mcitedefaultseppunct}\relax
\EndOfBibitem
\bibitem[Clemmer and Robbins(2022)]{Clemmer2022}
J.~T. Clemmer and M.~O. Robbins, \emph{Phys. Rev. Lett.}, 2022, \textbf{129},
  078002\relax
\mciteBstWouldAddEndPuncttrue
\mciteSetBstMidEndSepPunct{\mcitedefaultmidpunct}
{\mcitedefaultendpunct}{\mcitedefaultseppunct}\relax
\EndOfBibitem
\bibitem[Kirkwood(1939)]{Kirkwood1939}
J.~G. Kirkwood, \emph{J. Chem. Phys.}, 1939, \textbf{7}, 506--509\relax
\mciteBstWouldAddEndPuncttrue
\mciteSetBstMidEndSepPunct{\mcitedefaultmidpunct}
{\mcitedefaultendpunct}{\mcitedefaultseppunct}\relax
\EndOfBibitem
\bibitem[Schwartz \emph{et~al.}(1985)Schwartz, Feng, Thorpe, and
  Sen]{Schwartz1985}
L.~M. Schwartz, S.~Feng, M.~F. Thorpe and P.~N. Sen, \emph{Phys. Rev. B}, 1985,
  \textbf{32}, 4607--4617\relax
\mciteBstWouldAddEndPuncttrue
\mciteSetBstMidEndSepPunct{\mcitedefaultmidpunct}
{\mcitedefaultendpunct}{\mcitedefaultseppunct}\relax
\EndOfBibitem
\bibitem[Reid \emph{et~al.}(2018)Reid, Pashine, Wozniak, Jaeger, Liu, Nagel,
  and de~Pablo]{Reid2018}
D.~R. Reid, N.~Pashine, J.~M. Wozniak, H.~M. Jaeger, A.~J. Liu, S.~R. Nagel and
  J.~J. de~Pablo, \emph{Proc. Natl. Acad. Sci.}, 2018, \textbf{115},
  E1384--E1390\relax
\mciteBstWouldAddEndPuncttrue
\mciteSetBstMidEndSepPunct{\mcitedefaultmidpunct}
{\mcitedefaultendpunct}{\mcitedefaultseppunct}\relax
\EndOfBibitem
\bibitem[Clemmer and Robbins(2023)]{Clemmer2023b}
J.~T. Clemmer and M.~O. Robbins, 2023\relax
\mciteBstWouldAddEndPuncttrue
\mciteSetBstMidEndSepPunct{\mcitedefaultmidpunct}
{\mcitedefaultendpunct}{\mcitedefaultseppunct}\relax
\EndOfBibitem
\bibitem[Walton(1987)]{Walton1987}
K.~Walton, \emph{J. Mech. Phys. Solids}, 1987, \textbf{35}, 213--226\relax
\mciteBstWouldAddEndPuncttrue
\mciteSetBstMidEndSepPunct{\mcitedefaultmidpunct}
{\mcitedefaultendpunct}{\mcitedefaultseppunct}\relax
\EndOfBibitem
\bibitem[Greaves(2013)]{Greaves2013}
G.~N. Greaves, \emph{Notes Rec. R. Soc. J. Hist. Sci.}, 2013, \textbf{67},
  37--58\relax
\mciteBstWouldAddEndPuncttrue
\mciteSetBstMidEndSepPunct{\mcitedefaultmidpunct}
{\mcitedefaultendpunct}{\mcitedefaultseppunct}\relax
\EndOfBibitem
\bibitem[Leclerc(2019)]{Leclerc2019}
W.~Leclerc, \emph{Granul. Matter}, 2019, \textbf{21}, 17\relax
\mciteBstWouldAddEndPuncttrue
\mciteSetBstMidEndSepPunct{\mcitedefaultmidpunct}
{\mcitedefaultendpunct}{\mcitedefaultseppunct}\relax
\EndOfBibitem
\bibitem[Nguyen \emph{et~al.}(2019)Nguyen, Andr{\'{e}}, and Huger]{Nguyen2019}
T.-T. Nguyen, D.~Andr{\'{e}} and M.~Huger, \emph{Comput. Part. Mech.}, 2019,
  \textbf{6}, 393--409\relax
\mciteBstWouldAddEndPuncttrue
\mciteSetBstMidEndSepPunct{\mcitedefaultmidpunct}
{\mcitedefaultendpunct}{\mcitedefaultseppunct}\relax
\EndOfBibitem
\bibitem[Rojek \emph{et~al.}(2021)Rojek, Nosewicz, and Thoeni]{Rojek2021}
J.~Rojek, S.~Nosewicz and K.~Thoeni, \emph{Int. J. Numer. Methods Eng.}, 2021,
  \textbf{122}, 3335--3367\relax
\mciteBstWouldAddEndPuncttrue
\mciteSetBstMidEndSepPunct{\mcitedefaultmidpunct}
{\mcitedefaultendpunct}{\mcitedefaultseppunct}\relax
\EndOfBibitem
\bibitem[Groot and Warren(1997)]{Groot1997}
R.~D. Groot and P.~B. Warren, \emph{J. Chem. Phys.}, 1997, \textbf{107},
  4423--4435\relax
\mciteBstWouldAddEndPuncttrue
\mciteSetBstMidEndSepPunct{\mcitedefaultmidpunct}
{\mcitedefaultendpunct}{\mcitedefaultseppunct}\relax
\EndOfBibitem
\bibitem[Luan and Robbins(2005)]{Luan2005}
B.~Luan and M.~O. Robbins, \emph{Nature}, 2005, \textbf{435}, 929--932\relax
\mciteBstWouldAddEndPuncttrue
\mciteSetBstMidEndSepPunct{\mcitedefaultmidpunct}
{\mcitedefaultendpunct}{\mcitedefaultseppunct}\relax
\EndOfBibitem
\bibitem[Luan and Robbins(2006)]{Luan2006}
B.~Luan and M.~O. Robbins, \emph{Phys. Rev. E}, 2006, \textbf{74}, 1--17\relax
\mciteBstWouldAddEndPuncttrue
\mciteSetBstMidEndSepPunct{\mcitedefaultmidpunct}
{\mcitedefaultendpunct}{\mcitedefaultseppunct}\relax
\EndOfBibitem
\bibitem[Pastewka and Robbins(2016)]{Pastewka2016}
L.~Pastewka and M.~O. Robbins, \emph{App. Phys. Lett.}, 2016, \textbf{108},
  221601\relax
\mciteBstWouldAddEndPuncttrue
\mciteSetBstMidEndSepPunct{\mcitedefaultmidpunct}
{\mcitedefaultendpunct}{\mcitedefaultseppunct}\relax
\EndOfBibitem
\bibitem[Silbert \emph{et~al.}(2001)Silbert, Ertaş, Grest, Halsey, Levine, and
  Plimpton]{Silbert2001}
L.~E. Silbert, D.~Ertaş, G.~S. Grest, T.~C. Halsey, D.~Levine and S.~J.
  Plimpton, \emph{Phys. Rev. E}, 2001, \textbf{64}, 051302\relax
\mciteBstWouldAddEndPuncttrue
\mciteSetBstMidEndSepPunct{\mcitedefaultmidpunct}
{\mcitedefaultendpunct}{\mcitedefaultseppunct}\relax
\EndOfBibitem
\bibitem[Tsuji \emph{et~al.}(1992)Tsuji, Tanaka, and Ishida]{Tsuji1992}
Y.~Tsuji, T.~Tanaka and T.~Ishida, \emph{Powder Technol.}, 1992, \textbf{71},
  239--250\relax
\mciteBstWouldAddEndPuncttrue
\mciteSetBstMidEndSepPunct{\mcitedefaultmidpunct}
{\mcitedefaultendpunct}{\mcitedefaultseppunct}\relax
\EndOfBibitem
\bibitem[Mindlin(1949)]{Mindlin1949}
R.~D. Mindlin, \emph{J. Appl. Mech.}, 1949, \textbf{16}, 259--268\relax
\mciteBstWouldAddEndPuncttrue
\mciteSetBstMidEndSepPunct{\mcitedefaultmidpunct}
{\mcitedefaultendpunct}{\mcitedefaultseppunct}\relax
\EndOfBibitem
\bibitem[Luding(2008)]{Luding2008}
S.~Luding, \emph{Granul. Matter}, 2008, \textbf{10}, 235--246\relax
\mciteBstWouldAddEndPuncttrue
\mciteSetBstMidEndSepPunct{\mcitedefaultmidpunct}
{\mcitedefaultendpunct}{\mcitedefaultseppunct}\relax
\EndOfBibitem
\bibitem[Marshall(2009)]{Marshall2009}
J.~Marshall, \emph{J. Comput. Phys.}, 2009, \textbf{228}, 1541--1561\relax
\mciteBstWouldAddEndPuncttrue
\mciteSetBstMidEndSepPunct{\mcitedefaultmidpunct}
{\mcitedefaultendpunct}{\mcitedefaultseppunct}\relax
\EndOfBibitem
\bibitem[Clemmer \emph{et~al.}(2023)Clemmer, Long, and Brown]{Clemmer2023}
J.~T. Clemmer, K.~N. Long and J.~A. Brown, \emph{Mech. Mater.}, 2023,
  104693\relax
\mciteBstWouldAddEndPuncttrue
\mciteSetBstMidEndSepPunct{\mcitedefaultmidpunct}
{\mcitedefaultendpunct}{\mcitedefaultseppunct}\relax
\EndOfBibitem
\bibitem[Giannis \emph{et~al.}(2021)Giannis, Schilde, Finke, Kwade, Celigueta,
  Taghizadeh, and Luding]{Giannis2021}
K.~Giannis, C.~Schilde, J.~H. Finke, A.~Kwade, M.~A. Celigueta, K.~Taghizadeh
  and S.~Luding, \emph{Granul. Matter}, 2021, \textbf{23}, 17\relax
\mciteBstWouldAddEndPuncttrue
\mciteSetBstMidEndSepPunct{\mcitedefaultmidpunct}
{\mcitedefaultendpunct}{\mcitedefaultseppunct}\relax
\EndOfBibitem
\bibitem[O'Hern \emph{et~al.}(2003)O'Hern, Silbert, Liu, and Nagel]{OHern2003}
C.~S. O'Hern, L.~E. Silbert, A.~J. Liu and S.~R. Nagel, \emph{Phys. Rev. E},
  2003, \textbf{68}, 011306\relax
\mciteBstWouldAddEndPuncttrue
\mciteSetBstMidEndSepPunct{\mcitedefaultmidpunct}
{\mcitedefaultendpunct}{\mcitedefaultseppunct}\relax
\EndOfBibitem
\bibitem[Agnolin and Roux(2007)]{Agnolin2007}
I.~Agnolin and J.~N. Roux, \emph{Phys. Rev. E - Stat. Nonlinear, Soft Matter
  Phys.}, 2007, \textbf{76}, 1--22\relax
\mciteBstWouldAddEndPuncttrue
\mciteSetBstMidEndSepPunct{\mcitedefaultmidpunct}
{\mcitedefaultendpunct}{\mcitedefaultseppunct}\relax
\EndOfBibitem
\bibitem[Wang \emph{et~al.}(2021)Wang, Zhang, Tuckman, Ouellette, Shattuck, and
  O'Hern]{Wang2021b}
P.~Wang, S.~Zhang, P.~Tuckman, N.~T. Ouellette, M.~D. Shattuck and C.~S.
  O'Hern, \emph{Phys. Rev. E}, 2021, \textbf{103}, 22902\relax
\mciteBstWouldAddEndPuncttrue
\mciteSetBstMidEndSepPunct{\mcitedefaultmidpunct}
{\mcitedefaultendpunct}{\mcitedefaultseppunct}\relax
\EndOfBibitem
\bibitem[Carnavas and Page(1998)]{Carnavas1998}
P.~Carnavas and N.~Page, \emph{J. Mater. Sci.}, 1998, \textbf{33},
  4647--4655\relax
\mciteBstWouldAddEndPuncttrue
\mciteSetBstMidEndSepPunct{\mcitedefaultmidpunct}
{\mcitedefaultendpunct}{\mcitedefaultseppunct}\relax
\EndOfBibitem
\bibitem[Hentschel and Page(2007)]{Hentschel2007}
M.~L. Hentschel and N.~W. Page, \emph{J. Mater. Sci.}, 2007, \textbf{42},
  1261--1268\relax
\mciteBstWouldAddEndPuncttrue
\mciteSetBstMidEndSepPunct{\mcitedefaultmidpunct}
{\mcitedefaultendpunct}{\mcitedefaultseppunct}\relax
\EndOfBibitem
\bibitem[Patinet \emph{et~al.}(2016)Patinet, Vandembroucq, and
  Falk]{Patinet2016}
S.~Patinet, D.~Vandembroucq and M.~L. Falk, \emph{Phys. Rev. Lett.}, 2016,
  \textbf{117}, 045501\relax
\mciteBstWouldAddEndPuncttrue
\mciteSetBstMidEndSepPunct{\mcitedefaultmidpunct}
{\mcitedefaultendpunct}{\mcitedefaultseppunct}\relax
\EndOfBibitem
\end{mcitethebibliography}
\bibliographystyle{rsc}

\end{document}